\documentclass[11pt,onecolumn,floatfix,superscriptaddress]{revtex4}

\usepackage{graphicx}                   
\usepackage{epsfig}
\usepackage{amsmath,amssymb}            
\usepackage{hyperref}                   
\usepackage{epstopdf}
\usepackage{placeins}
\usepackage{upgreek}
\usepackage{slashed}
\usepackage{color}
\definecolor{orange}{rgb}{1,0.5,0}
\definecolor{brown}{rgb}{0.65, 0.16, 0.16}
\definecolor{phlox}{rgb}{0.87, 0.0, 1.0}

\graphicspath{{figs/}}                  

\begin{document}

    \title{Cosmic Birefringence as a probe of dark matter nature: Sterile neutrino and dipolar dark matter }
\author{Jafar Khodagholizadeh}
\email{gholizadeh@ipm.ir}
\affiliation{Farhangian University, P.O. Box 11876-13311, Tehran, Iran}
\author{S. Mahmoudi}
\email{s.mahmoudi@shirazu.ac.ir}
\affiliation{ Department of Physics, School of Science, Shiraz University, Shiraz 71454, Iran }
\author{ Rohoollah Mohammadi}
\email{rmohammadi@ipm.ir}
\affiliation{Iranian National Museum of Science and Technology (INMOST), PO BOX: 11369-14611, Tehran, Iran,\\
	School of Astronomy, Institute for Research in Fundamental Sciences (IPM), P. O. Box 19395-5531, Tehran, Iran}
\author{ Mahdi Sadegh}
\email{m.sadegh@ipm.ir}
\affiliation{School of Particles and Accelerators, Institute for Research in Fundamental
	Sciences (IPM), P. O. Box 19395-5531, Tehran, Iran.}

    \begin{abstract}
    Recently, non-zero rotation angle $\beta=0.30^\circ\pm0.11^\circ$ $(68\%\text{ C.L.})$ [Phys. Rev. Lett. \textbf{128}, no.9, 091302 (2022)] has been reported for linear polarization of cosmic microwave background (CMB) radiation, which is known as cosmic birefringence (CB). We used this birefringence angle of CMB to study and distinguish different candidates of dark matter (DM), e.g., dipolar and sterile neutrino DM. We calculated CMB forward scattering by those probable candidates of DM to generate $\beta$ in the presence of primordial scalar fluctuations' background. We explicitly plotted bounds on the mass and electromagnetic coupling for different sectors of DM, sterile neutrino, and dipolar DM, and compared them with other experimental bounds. Regarding dipolar DM, our calculations put a bound on the Majorana magnetic dipole moment about $\mathcal{M}\leqslant 1.4\times10^{-14}\,\frac{\beta}{0.30^\circ}\sqrt{\frac{m_{\text{\tiny{DM}}}}{1\,GeV}}\, e.\text{\,cm}$. In the case of sterile neutrino DM, the bound on the mass and mixing angle was estimated at $\theta^2 \leqslant 3.3\,(rad)^2\frac{\beta}{0.30^\circ}\,\frac{m_{DM}}{\rm{KeV}} $, which can be a new constraint for sterile neutrino DM whose production mechanism is motivated by models with a hidden sector coupled to the sterile neutrino. Based on our results, if the constraint on the mass and the electromagnetic coupling for DM must be within the allowed region, none of the considered candidates can compensate for all the observed CB angle. We also discussed the maximum contribution of the CB angle via CMB forward scattering by different sectors of the dark matter.

    \end{abstract}


    \maketitle

   \section{Introduction}
   Astrophysical observations of the past several decades confirm that the majority of matter in the universe consists of DM \cite{rotation, White:1977jf, WMAP:2003elm, Clowe:2003tk}. Up to date, several attempts have been made to identify the physical properties of the missing mass, including accelerator-based techniques \cite{Cai:2017mow, Baer:2012uy, Sigurdson:2004zp, Kumar:2015wya}, direct detection \cite{Catena:2018uae, Graham:2013gfa, Duerr:2015aka}, indirect search experiments \cite{Arguelles:2012cf, Foster:2021fxa, Sigurdson:2004zp, Ramos:2021omo}, and astroparticle approaches such as studying their possible effects on the CMB radiation \cite{Mahmoudi:2018lll, ModaresVamegh:2019nja, Haghighat:2019rht}. 
 However, despite significant endeavors to determine its nature and properties, the identity of DM is still shrouded in mystery in astronomy and particle physics \cite{Green:2021jrr, Cooley:2021rws, Slatyer:2021qgc}. \par
 It seems that DM consists of non-relativistic particles that mainly interact gravitationally. In addition to gravitational interactions, they may experience very weak interaction with the standard model. Regarding the electromagnetic interaction, their coupling to photons is believed to be either non-existent or feeble. Nevertheless, an anomalous coupling with photons has been proposed for some candidates of DM which causes various phenomena in astrophysics and cosmology, such as superradiance around rotating black holes \cite{Arvanitaki:2014wva, Cardoso:2018tly, Stott:2018opm}, X- or $\gamma$-ray emission from the decay of DM \cite{Hofmann:2019ihc, Chiang:2014xra, Cadamuro:2011fd, Arias:2012az, Higaki:2014qua}, and cosmic birefringence (CB) \cite{Pospelov:2008gg, Fedderke:2019ajk, Agrawal:2019lkr}.\par
 
 The CB angle can be generated in parity violating interaction \cite{Lue:1998mq, Fujita:2020ecn, Liu:2006uh}. This angle measures the possible rotation of the linear polarization plane for photons propagating over large distances in the universe. Among all cosmic rays, CMB photons are an ideal target to probe this effect for the following reasons: They are emitted at the epoch of recombination, accurate prediction of their polarization angular power spectra by the $\Lambda$ cold DM ($\Lambda$CDM) model is possible, and the CMB polarization is sensitive to parity-violating phenomena. Indeed, it is conventional to decompose the observed pattern of CMB polarization into eigenstates of parity referred to as E and B modes whose correlations, which are zero in the standard scenario of cosmology due to parity conservation, lead to CMB birefringence. \par
 
 Generally, CB can be considered a probe of physics beyond the standard model of cosmology and elementary particles which breaks Lorentz and CPT symmetry. For instance, it provides a tantalizing hint for new physics of axions. The interaction between CMB photons and axion-like particles (ALPs) with various potentials occurring during or after the recombination epoch could be accounted for isotropic CB. For the quadratic and cosine potentials, the lower bounds on some physical properties of APL have been obtained using the observational value of CB \cite{Fujita:2020ecn}. An axion with a linear potential that plays the role of dark energy has been studied in \cite{Gasparotto:2022uqo}, where an upper bound has been placed on the axion decay constant using the birefringence measurement and the constraint on the equation of state for dark energy by Planck 2018 \cite{Minami:2020odp} and the axion-photon coupling constant by $Chandra$.\par
 
 There are several independent approaches for extracting the birefringence angle; one such approach on which most discussions of birefringence have focused is using $EE$, $BB$ and $EB$ power spectra \cite{Minami:2020odp, WMAP:2008lyn, Planck:2015qep, Namikawa:2020ffr, ACT:2020frw, Komatsu:2022nvu}. However, this approach has a fundamental problem. Since this effect degenerates with an artificial rotation of polarization angles generated by orientation miscalibration of polarimeters, it is not possible to distinguish birefringence from the systematic uncertainty of a miscalibration of the orientation angle. The galactic foreground emission is one way to deal with this problem \cite{Minami:2020fin}. To this end, given that the miscalibration of detector orientation changes the foreground and the CMB spectra in the same way, one could use data at different frequencies to separate birefringence from foreground and calibration effects. Another independent method has been proposed by authors in \cite{Abghari:2022bet}, based on which the birefringence angle is determined only by using CMB temperature, $E$ modes, and their cross-correlation. In addition to the above methods, it has recently been found that parity violating forward scattering of the CMB in the presence of scalar perturbation can lead to the CB effect. For instance, the authors in \cite{Mohammadi:2021xoh} have shown that the weak interaction of the CMB and cosmic neutrino background, in the order of one loop forward scattering, can generate non-vanishing CB in the presence of scalar perturbation whose value is at least one order larger than the CB angle reported by using Planck data release. Therefore, other mechanisms resulting in non-vanishing parity violating forward scattering should be taken into consideration to produce the CB \cite{Khodagholizadeh:2014nfa,Mohammadi:2016bxl, Khodagholizadeh:2019het,Tizchang:2016vef,2022dipole}. \par
 
 Following the last mentioned approach, here, we introduce the two new sources of CB, i.e., dipolar DM and sterile neutrino DM. We show that the birefringence limit from CMB observations enables us to constrain their coupling to photons. In other words, the CB effect provides a new tool to investigate the DM properties and open up a new observational window to explore the nature of DM.\par
 
 The rest of the paper is organized as follows: We present a brief review of relativistic Boltzmann equations in section \ref{Sec:method}. In section \ref{CB}, we give a general discussion of CB. Then, we choose dipolar DM and sterile neutrino DM as two kinds of DM and study their interaction effects on the linear polarization of the CMB which result in CB in section \ref{DM}. In section \ref{conclusion}, we present the summary and conclusion.

  \section{Relativistic Boltzmann Equations }\label{Sec:method}

 The linear and circular polarizations of an ensemble of photons described by Stokes parameters are the density matrix components in the polarization space as follows:
   \begin{eqnarray}\label{eq:rho}
  	\hat{\rho}_{ij}\equiv\frac{1}{2}\left(
  	\begin{matrix}
  		I+Q & U-iV \\
  		U+iV&  I-Q \\
  	\end{matrix}
  	\right),
  \end{eqnarray}
  \color{black}
  where $I$ is the total intensity of radiation, $U$, $Q$, and $V$ describe the polarization intensity of photons and, for unpolarized photons, $Q=U=V=0$. The linear polarization of the photon is defined in terms of the Stokes parameters $Q$ and $U$, and parameter $V$ indicates the net circular polarization or the difference between left- and right-circular polarization intensities. \par
 The time evolution of the density matrix components $\rho_{ij}({\bf{k}})$ is given by the quantum Boltzmann equation as \cite{kosowsky1996cosmic}
  \begin{eqnarray}\label{forward}
  (2\pi)^3\delta^3(0)(2k^0)\dfrac{d}{dt}\rho_{ij}({\bf{k}})=i\langle[H_{I}^0(t);D_{ij}^{0}({\bf{k}})]\rangle -\dfrac{1}{2}\int dt \langle[H_{I}^0(t);[H_{I}^0(t);D_{ij}^{0}({\bf{k}})]]\rangle,
  \end{eqnarray}
    where $H_{I}^0(t)$ is the first order of the interacting Hamiltonian
  \begin{equation}\label{HI}
  	H_{I}^0(t)=-\frac{i}{2}\int\limits_{-\infty}^{\infty}\,dt'T\{H(t),H(t')\},
  \end{equation}
 in which $T$ signifies a time-ordered product and $H$ denotes the interaction Hamiltonian which relates to the interaction Hamiltonian density $\mathcal{H}(x)$ as follows
  \begin{equation}
  	H(t)=\int d^{3}{\bf{x}}\,\mathcal{H}(x).
  \end{equation}
    \color{black}
   Moreover, $ D_{ij} ({\bf{k}})\equiv a_{i}^{\dagger}({\bf{k}}) a_{i}({\bf{k}}) $ is the photon number operator. The first term on the right-hand side of Eq. \eqref{forward} is called the forward scattering term which is proportional to the amplitude of the scattering, whereas the second one is known as the higher order collision term, giving the scattering cross section which is highly subdominant compared to the first term.\par
 
 Although the Stokes parameters $I$ and $V$ are independent of the reference frame, $Q$ and $U$ are frame-dependent parameters. However, by introducing a set of linear combinations of polarization parameters $Q$ and $U$ as $\Delta^{\pm}_{\text{P}}=Q\pm i U$, one can find the reference frame-independent parameters, where $P$ stands for polarization.
 Accounting for the scalar mode perturbation of the metric and using the Thomson scattering of CMB photons by cosmic electrons, the time evolution of the CMB radiation transfer is governed by the following set of equations \cite{Bond:1984fp}:
\begin{eqnarray}
	&&\dfrac{d}{d\eta}\Delta^{\text{S}}_{\text{I}}+iK\mu\Delta^{\text{S}}_{\text{I}} +4[\dot{\varPsi}-iK\mu\Phi]=\dot{\tau_{\text{e}}}\left[ -\Delta^{\text{S}}_{\text{I}}+\Delta^{\text{S}}_{\text{I}_{\circ}}+i\mu v_{b}+\dfrac{1}{2}P_{2}(\mu)\Pi\right],\\
	&&\dfrac{d}{d\eta}\Delta^{\pm \text{S}}_{\text{P}}+iK\mu\Delta^{\pm \text{S}}_{\text{P}} =\dot{\tau_{\text{e}}}\left[ -\Delta^{\pm \text{S}}_{\text{P}}-\dfrac{1}{2}[1-P_{2}(\mu)]\Pi \right],
\end{eqnarray}
 where $\Psi$ and $\Phi$ are the metric perturbations, $v_{b}$ is the baryon bulk velocity, and
 \begin{equation}
 	\Pi\equiv \Delta_{I2}^{(\rm S)} +\Delta_{P2}^{(\rm S)}+\Delta_{P0}^{(\rm S)}
 \end{equation}
  The superscript ``S" denotes the scalar metric perturbations, and the CMB radiation transfer is described by the multipole moments of temperature $(\text{I})$ and polarization $(\text{P})$ \cite{Zaldarriaga:1996xe, Seljak:1996is}
 \begin{equation}\label{BE1}
 \Delta^{\text{S}}_{\text{I,P}}(\eta,K,\mu)=\sum_{l=0}^{\infty}(2l+1)(-i)^l\Delta^{\text{S}}_{\text{I,P}_{l}}(\eta,K)P_{l} (\mu),
 \end{equation}
where $P_{l}(\mu)$ is the Legendre polynomial of rank $l$, $\mu=\hat{\bf{n}}.\hat{K}= \cos\theta$, and $\theta$ is the angle between the CMB photon direction $\hat{\bf{n}}=\frac{\bf{k}}{|\bf{k}|}$ and the wave vectors $K$ of the Fourier modes of scalar perturbations (S).
 Besides, $\dot{\tau}_{e}=an_ex_e\sigma_T$ indicates the differential optical depth for Thomson scattering in which $\sigma_T$ is the Thomson cross-section and $a(\eta)$ is the scale factor normalized to unity at present as a function of conformal time ($\eta$). Electron density and ionization fraction are denoted by $n_e$ and $x_e$, respectively. \par 
 To study the polarization features of the CMB photons in the context of cosmology, it is common to separate the CMB polarization $\Delta^{\pm\text{S}}_{\text{P}}(\eta,K,\mu)$ into a curl-free part (E mode) and a divergence-free part (B mode) as follows:
   \begin{eqnarray}
   \Delta^{\text{S}}_{\text{E}}(\eta_{0},K,\mu)&\equiv&-\dfrac{1}{2}\left[\bar{\eth}^{2}\,\Delta^{+\text{S}}_{\text{P}}(\eta_{0},K,\mu)+\,\eth^{2}\Delta^{-\text{S}}_{\text{P}}(\eta_{0},K,\mu) \right],\label{BE11}\\
   \Delta^{\text{S}}_{\text{B}}(\eta_{0},K,\mu)&\equiv&\dfrac{i}{2}\left[\bar{\eth}^{2}\,\Delta^{+\text{S}}_{\text{P}}(\eta_{0},K,\mu)-\eth^{2}\Delta^{-\text{S}}_{\text{P}}(\eta_{\circ},K,\mu) \right],\label{BE12}
   \end{eqnarray}
  in which $\eth $ and $\bar{\eth}$ are spin raising and lowering operators, respectively. One can obtain the value of $\Delta _{E,B}^{S}(\hat{\bf n})$ at the present time $\eta_0$ and in the direction $\bf \hat{n}$ by summing over all their Fourier modes $K$ as follows \cite{Zaldarriaga:1996xe, Seljak:1996is}:
   \begin{eqnarray}
   \Delta _{E,B}^{S}(\hat{\bf{n}})
   &=&\int d^3 {\bf K} \,\xi({\bf K})e^{\mp2i\phi_{K,n}}\Delta _{E,B}^{S}
   (\eta_0,K,\mu),\label{sumf}
   \end{eqnarray}
  where $\phi_{K,n}$ is the angle needed to rotate the $\bf{K}$ and $\hat{\bf{n}}$ dependent basis to a fixed frame in the sky. Moreover, it should be mentioned that the negative sign is used for the E mode and the positive sign is for the B mode The random variable $\xi(\bf{K})$ used to characterize the initial amplitude of the mode satisfies
   \begin{equation}
   \langle \xi^{*}({\bf K}_1)\xi({\bf K}_2)
   \rangle=
   P_{S}({\bf K})\delta({\bf K}_1- {\bf K}_2),\label{sps}
   \end{equation}
  where $P_S(K)$ is the initial power spectrum of the scalar mode perturbation and the angle brackets $\langle...\rangle$ represent an ensemble average over initial conditions. Finally,
   to characterize the statistics of the CMB perturbations we need to calculate the power spectra which is defined as the rotationally invariant quantity as follows
  \begin{equation}\label{cll}
  	C_{E,B}^{\ell}=\frac{1}{2\ell+1}\sum_{m}\left\langle a^{\ast}_{E,lm} a_{B,lm} \right\rangle,
  \end{equation}
  where $a_{E,lm}$ and $a_{B,lm}$ are the expansion coefficients of $\Delta _{E,B}^{S}(\hat{\bf n})$ in terms of spherical
harmonics
  \begin{eqnarray}
  \Delta _{E}^{S}(\hat{{\bf n}})
  &=&\sum_{lm}\left[{(l+2)! \over (l-2)!}\right]^{1/2}
  a_{E,lm}Y_{lm}(\hat{{\bf n}}), \nonumber \\  
  \Delta _{B}^{S}(\hat{\bf n})
  &=&\sum_{lm}\left[{(l+2)! \over (l-2)!}\right]^{1/2}
  a_{B,lm}Y_{lm}(\hat{\bf n}). 
  \label{EB} 
  \end{eqnarray}
  Now, using Eq.\eqref{cll} and by integrating Eqs.~(\ref{sumf}) and (\ref{sps}) over the initial power spectrum of the metric perturbation, we obtain the power spectrum for $E$- and $B$- modes as follows
   \begin{equation}\label{Cl1}
   C_{E,B}^{\ell\,(S)}=\frac{1}{2\ell+1}\frac{(\ell-2)!}{(\ell+2)!}\int d^3{\bf{K}} P_S({\bf{K}})\Big|\sum_m\int d\Omega\, Y^*_{lm}(\hat{{\bf n}})\Delta_{E,B}^{S}(\eta_0,K,\mu)\Big|^2.
   \end{equation}
 
    \color{black}
\section{General discussion about Cosmic Birefringence (CB)}\label{CB}
 In the context of the standard model of cosmology, due to the parity symmetry, $E$ and $B$ mode polarizations are not correlated. However, if CMB photons experience some sort of interaction that violates parity symmetry and the Lorentz symmetry, the phase velocities of the right- and left-hand helicity states of photons will differ and, therefore, the plane of linear polarization will rotate in the sky by an angle $\beta$ 
        \begin{equation}
    	\label{eqn:rotation}
    	Q\pm iU\mapsto\left(Q\pm iU\right)e^{\pm2i\beta},
    \end{equation}
 where the function $\beta$, called \emph{birefringence angle}, characterizes the amplitude of deviation from the standard model. Hence, a part of $E$-modes' polarization transfers into $B$-mode ones, and the $EB$ mode polarization will not be zero. 
 Such an effect could have left measurable imprints in the CMB angular power spectra $C_{\ell}$'s. Indeed, the presence of parity violation interaction induces a rotation of the CMB angular power spectra as follows:
   \begin{eqnarray}
  C_{EE,\ell}&=&\cos^2(2\beta)\bar{C}_{EE,\ell}+\sin^2(2\beta)\bar{C}_{BB,\ell},\nonumber\\
  C_{BB,\ell}&=&\cos^2(2\beta)\bar{C}_{BB,\ell}+\sin^2(2\beta)\bar{C}_{EE,\ell},\\
  C_{EB,\ell}&=&\dfrac{1}{2}\sin(4\beta)(\bar{C}_{EE,\ell}-\bar{C}_{BB,\ell}),\nonumber
  \end{eqnarray} 
 where $\bar{C}_{EE,\ell}$ is the standard $E$-mode power spectra, and $\beta=0.30^\circ\pm0.11^\circ$ ($68\%\text{C.L.}$) is the value of the CB angle reported by using Planck data release \cite{Diego-Palazuelos:2022dsq}. As the above equations show, in the presence of parity-violating interaction, space acts like a birefringent material, and the cross-correlators EB turn on. In contrast, by setting $\beta=0$, one can recover the standard results of the CMB angular power spectra.\par
    \section{Dark matter's  impact on the linear polarization}\label{DM}
Regarding the possible features of DM, various experiments based on direct or indirect methods of detection have been proposed to explore its properties. Here, we use the CB effect as a way to indirectly examine the DM signatures. The key point is that if the physics behind DM leads to the parity symmetry violation assumed in the standard model of cosmology, the linear polarization of the CMB can rotate due to its coupling to this dark sector. This results in a non-vanishing B mode and parity-violating EB correlations. Taking into account this point, we consider two kinds of DM, i.e., dipolar DM and sterile neutrino DM, whose coupling to the photon would induce a rotation of the polarization plane of the CMB, and try to study their properties through CMB birefringence.
   
  \subsection{Dipolar dark matter}
 The effective Lagrangian for the coupling of the electromagnetic field $F^{\mu\nu}$ with a Dirac fermion that possesses a magnetic dipole moment $\mathcal{M}$ and an electric dipole moment $\mathcal{D}$ is as follows:
 \begin{equation}\label{DDM}
 \mathcal{L}_{\rm DDM}=-\frac{i}{2}\bar{\psi}\sigma_{\rm \mu\nu}(\mathcal{M}+\gamma^5 \mathcal{D})\psi F^{\mu\nu},
 \end{equation}
  where $\sigma^{\mu\nu}$ is the commutator of two Dirac matrices, $\sigma^{\mu\nu}=\frac{i}{2}[\gamma^\mu,\gamma^\nu]$. Regarding Majorana fermions, only non-zero transition multipole moments between different mass eigenstates can be defined, and their interaction with photons is described by \cite{Masso:2009mu, Lee:2014koa}
  \begin{equation}\label{DDM2}
  \mathcal{L}_{\rm DDM}=-\frac{i}{2}\bar{\psi}_{2}\sigma_{\rm \mu\nu}(\mathcal{M}_{12}+\gamma^5 \mathcal{D}_{12})\psi_{1} F^{\mu\nu} + H.c.,
  \end{equation}
 where $\mathcal{M}_{12}$ is a transition magnetic moment, and $\mathcal{D}_{12}$ is a transition electric moment. 
 This possible interaction between photon and dipolar DM opens up a new window to explore the features of this candidate of DM. Here, we focus on the Majorana DM which interacts with CMB photons via transition magnetic dipole moment and study its properties using the CB effect of the CMB. To this end, we consider two singlet Majorana fermions $\chi_{1}$ and $\chi_{2}$, with mass splitting $\delta$, which couple to photons by the magnetic dipole interaction Lagrangian \cite{Davidson:2005cs}
  \begin{equation}\label{majoranaDM}
  \mathcal{L}_{\text{\tiny{DDM}}}=-\frac{i}{2}\,\mathcal{M}_{12}\,\bar{\chi}_{1}\sigma_{\rm \mu\nu}\chi_{2} F^{\mu\nu} + H.C..
  \end{equation}
  Assuming the singlet Majorana fermions to be the right-handed neutrinos $(\chi\,=\,\psi_{R}+\psi_{R}^{c})$, the Lagrangian (\ref{majoranaDM}) casts into 
  \begin{equation}\label{majorana}
  	\mathcal{L}_{\text{\tiny{DDM}}}=-\frac{i}{2} \,\,\mathcal{M}_{12}\,(\,\bar{\psi^{c}}_{1}\sigma_{\rm \mu\nu} P_{R}\psi_{2}\,\,F^{\mu\nu}\,\,+\,\,\bar{\psi}_{1}\sigma_{\rm \mu\nu} P_{L}\psi^{c}_{2}\,\,F^{\mu\nu})+H.C., 
  \end{equation}
  where $P_{R}=\frac{1}{2}(1+\gamma^{5})$, $P_{L}=\frac{1}{2}(1-\gamma^{5})$ and $\psi^{c}=-i\gamma_{2}\psi^{\star}$. \par
\begin{figure}
 	\includegraphics[scale=0.8]{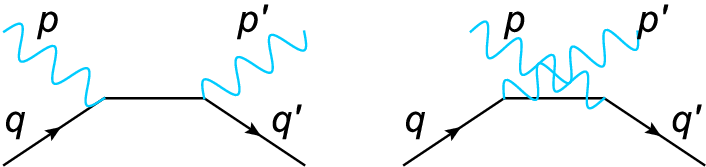}\\
 	\caption{The typical diagrams for photon-DM scattering}\label{fig1}
 \end{figure}

 The Feynman diagram corresponding to this interaction at the lowest order is shown in Fig.\ref{fig1}. 
To investigate the possible effects of this interaction on the CMB polarization at the forward scattering level, one needs to calculate $ \langle [H_{I}^0(0), D_{ij}({\bf{k}})]\rangle$ which is obtained as \cite{Mahmoudi:2018lll}
  \begin{equation}\label{DDM25}
  	i\langle [H_{I}^0(0), D_{ij}({\bf{k}})]\rangle = i\int d{\bf{q}}\,n_{\text{\tiny{DM}}}({\bf{x}},{\bf{q}})(\delta_{is}\rho_{s'j}({\bf{k}})-\delta_{js'}\rho_{is}({\bf{k}}))(2\pi)^3 \delta^{(3)}(0){M}\,\,\vert_{{\tiny{q'=q,\,\,p'=p=k, r=r'}}},
  \end{equation}	
   where ${d \bf{q}} \equiv \frac {d^3 \bf{q}}{(2\pi)^3}\frac{m_{\text{{\tiny{DM}}}}}{q^0}$, $\rho_{ss'}({\bf{k}})$ indicates the elements of the CMB photon density matrix and $n_{\text{\tiny{DM}}}({\bf{x}},{\bf{q}})$ is the number density of DM. Moreover, ${M}$ denotes the total Feynman amplitude given by the following equation in the non-relativistic limit:
   \begin{eqnarray}\label{majorana122}
   	{M}&=& -\mathcal{M}_{12}^{2}\frac{(2k.q)^{2}}{(2k.q)^{2}-(m_{\text{\tiny{DM}}_{2}}^{2}-m_{\text{\tiny{DM}}_{1}}^{2})^{2}}\Big({\bf{k}}\cdot(\vec{\epsilon}_{s'}\times\vec{\epsilon}_{s})-k^{0}v(\vec{\epsilon}_{s'}\times\vec{\epsilon}_{s})\cdot{\hat{{\bf{v}}}}\Big)+(1\leftrightarrow2),
   \end{eqnarray}
   where $v=|\vec{q}|/m_{\text{\tiny{DM}}}$ denotes the velocity of DM and $\vec{\epsilon}_{s}(k)$'s are the photon polarization vectors with $s=1,2$ for two physical transverse polarization of a free photon.
 Considering the case that $\delta=m_{\text{\tiny{DM}}_{2}}-m_{\text{\tiny{DM}}_{1}}\ll k^0$, (\ref{majorana122}) can be estimated as follows
  \begin{equation}\label{DDM23}
  	{M}\simeq - \mathcal{M}_{12}^2\,\,{\bf{k}}\cdot(\vec{\epsilon}_{s'}\times\vec{\epsilon}_{s})\,+\,(1\,\leftrightarrow\,2),
  \end{equation}
  where since the order of the second term is smaller than
  the first one due to the presence of $v$, we ignore
  the terms proportional to the DM velocity. Moreover, it is important to note that in the case $\delta\gg k^0$,  the contribution of the
photon-dipolar DM scattering on the CMB polarization will be suppressed as $(\frac{k^0}{\delta})^2$ compared to the case $\delta\ll k^0$ and therefore, we will not consider this case for the rest of the paper (see appendix A for more details).
 
    By substituting (\ref{DDM23}) in (\ref{DDM25}) and using (\ref{forward}), the time evolution of the density matrix element can be written as
  \begin{equation}\label{DDM244}
  	\frac{d\rho_{ij}}{dt}=- i \mathcal{M}^2\,\,\big(n_{\text{\tiny{DM}}_{1}}({\bf{x}})+n_{\text{\tiny{DM}}_{2}}({\bf{x}})\big)(\delta_{is}\rho_{s'j}({\bf{k}})-\delta_{js'}\rho_{is}({\bf{k}}))(\vec{\epsilon}_{s'}\times\vec{\epsilon}_{s})\cdot \hat{{\bf{k}}},
  \end{equation}
  where $\mathcal{M}_{12}^{2}=\mathcal{M}_{21}^{2}=\mathcal{M}^{2}$, $\hat{{\bf{k}}}={\bf{k}}/k^0$ and the DM number density $n_{\text{\tiny{DM}}_{\text{\tiny{i}}}}$ ($\text{i}=1,2$) is
  \begin{equation}\label{DDM29}
  	n_{\text{\tiny{DM}}_{\text{\tiny{i}}}}({\bf{x}})=\int \frac{d^3 {\bf{q}}}{(2\pi)^3}\,\,n_{\text{\tiny{DM}}_{\text{\tiny{i}}}}({\bf{x}},{\bf{q}}).	
  \end{equation}
   Since $n_{\text{\tiny{DM}}_{1}}({\bf{x}})+n_{\text{\tiny{DM}}_{2}}({\bf{x}})=n_{\text{\tiny{DM}}}({\bf{x}})$, (\ref{DDM244}) is reduced to
  \begin{equation}\label{DDM2444}
  	\frac{d\rho_{ij}}{dt}=- i \mathcal{M}^2\,\,n_{\text{\tiny{DM}}}({\bf{x}})\,\,(\delta_{is}\rho_{s'j}({\bf{k}})-\delta_{js'}\rho_{is}({\bf{k}}))(\vec{\epsilon}_{s'}\times\vec{\epsilon}_{s})\cdot \hat{{\bf{k}}},
  \end{equation}
 and, consequently, the Stokes parameters evolve as
  \begin{eqnarray}
  	\frac{d I}{dt}&=& C_{e\gamma}^I,\label{inten}\\
  	\frac{d}{dt}(Q\pm iU)&=& C_{e\gamma}^\pm \mp i\dot{\tau}_{\text{\tiny{DM}}}(Q\pm iU),\label{Linpol}\\
  		\frac{d V}{dt}&=& C_{e\gamma}^V,\label{circpol}
  \end{eqnarray}
    where $C_{e\gamma}^I$,$C_{e\gamma}^V$ and $C_{e\gamma}^\pm$ show the contribution of Thomson scattering \cite{Kosowsky:1994cy} and $\dot{\tau}_{\text{DM}}$ is defined as follows:
  \begin{equation}\label{DDM3}
  \dot{\tau}_{\text{\tiny{DM}}}=\frac{ \mathcal{M}^2}{m_{\text{\tiny{DM}}}}\, \rho_{\text{\tiny{DM}}} ,
  \end{equation}
 in which $\rho_{\text{\tiny{DM}}}$ is the DM mass density. The second term in the right-hand side of \eqref{Linpol} shows that the photon-dipolar DM forward scattering affects the time evolution of the linear polarization of the CMB
      \begin{equation}\label{BE3}
  \dfrac{d}{d\eta}\Delta^{\pm \text{S}}_{\text{P}}+iK\mu\Delta^{\pm \text{S}}_{\text{P}} =\dot{\tau_{\text{e}}}\left[ -\Delta^{\pm \text{S}}_{\text{P}}-\dfrac{1}{2}[1-P_{2}(\mu)]\Pi \right]\mp i a(\eta)\dot{\tau}_{\text{\tiny{DM}}}\Delta^{\pm \text{S}}_{\text{P}},
  \end{equation}
  which leads to the following equation of polarization anisotropy:
    \begin{equation}\label{BE5}
  \dfrac{d}{d\eta}\left[ \Delta^{\pm \text{S}}_{\text{P}}e^{iK\mu\eta\pm i\tilde{\tau}_{\text{\tiny{DM}}}+\tilde{\tau}_{\text{e}}} \right]=-\dfrac{1}{2}e^{iK\mu\eta\pm i\tilde{\tau}_{\text{\tiny{DM}}}+\tilde{\tau}_{\text{e}}}\dot{\tau_{\text{e}}}[1-P_{2}(\mu)]\Pi,
  \end{equation}
  where
  \begin{equation}\label{BE6}
  \tilde{\tau}_{\text{\tiny{DM}}}(\eta,\mu)\equiv\int_{0}^{\eta}d \eta a(\eta)\dot{\tau}_{\text{\tiny{DM}}},\,\,\,\,\,\,\,\,\,\,\,\,\tilde{\tau}_{\text{e}}(\eta)\equiv\int_{0}^{\eta}d\eta\,a(\eta)\,\dot{\tau}_{\text{e}}.
  \end{equation}
  After calculation of $\Delta^{\pm \text{S}}_{P}$ and using \eqref{BE11} and \eqref{BE12}, the E-mode and B-mode polarizations produced through the dipolar DM-photon interaction will be obtained as follows:
    \begin{eqnarray}
    	\Delta^{\text{S}}_{\text{E}}(\eta_{0},K,\mu)&=&-\dfrac{3}{4}\int_{0}^{\eta_{0}} d\eta\,g_{\text{e}}(\eta)\Pi(\eta,K)\partial^{2}_{\mu}[(1-\mu^{2})e^{ix\mu}\cos\tau_{\text{\tiny{DM}}}],\label{BE14}\\
    	\Delta^{\text{S}}_{\text{B}}(\eta_{0},K,\mu)&=&\dfrac{3}{4}\int_{0}^{\eta_{0}} d\eta\,g_{\text{e}}(\eta)\Pi(\eta,K)\partial^{2}_{\mu}[(1-\mu^{2})e^{ix\mu}\sin\tau_{\text{\tiny{DM}}}],\label{BE15}
    \end{eqnarray}
    where $x=K(\eta_{0}-\eta)$ and $g_{\text{e}}(\eta)\equiv\dot{\tau}_{\text{e}} e^{-\tau_{\text{e}}}$ is the visibility function of the electron and describes the probability
    	that a photon scattered at epoch $\eta$ reaches the observer at
    	the present time, $\eta_0$ \cite{Dodelson:2003ft}. Finally, the power spectrum of the E and B-modes will be obtained by integrating over the initial power spectrum of the metric perturbation as:
    \begin{eqnarray}
  C_{EE,\ell}^{(\rm S)}\Big|_{\rm DM}&=& (4\pi)^{2} \frac{(\ell+2)!}{(\ell-2)!} \int d^{3}K P_{S}(K) \left[ \dfrac{3}{4}\int_{0}^{\eta_{0}}d\eta \, g_{e}(\eta)\, \Pi(K,\eta)\, \dfrac{j_{\ell}(x)}{x^{2}} \cos (\tau_{DM})\right]^2, \label{ClEE}
  \end{eqnarray}
  \begin{eqnarray}
  C_{BB,\ell}^{(\rm S)}\Big|_{\rm DM}&=& (4\pi)^{2} \frac{(\ell+2)!}{(\ell-2)!} \int d^{3}K P_{S}(K) \left[ \dfrac{3}{4}\int_{0}^{\eta_{0}}d\eta \, g_{e}(\eta)\, \Pi(K,\eta)\, \dfrac{j_{\ell}(x)}{x^{2}} \sin (\tau_{DM})\right]^2. \label{ClBB}
  \end{eqnarray}
  Moreover, the cross-power spectrum will be as follows:
  \begin{eqnarray}\label{12}
  C_{EB,\ell}^{(\rm S)}\Big|_{\rm DM}&=& \dfrac{(4\pi)^{2} }{4}\frac{(\ell+2)!}{(\ell-2)!} \int d^{3}K P_{S}(K) \left[ \dfrac{3}{4}\int_{0}^{\eta_{0}}d\eta \, g_{e}(\eta)\, \Pi(K,\eta)\, \dfrac{j_{\ell}(x)}{x^{2}}\sin(2\tau_{DM})\right]^2. \label{ClEB}
  \end{eqnarray}
 Since the standard E-mode coming from the Compton scattering in the CMB, i.e., $ \bar{C}_{EE,\ell}^{(\rm S)} $, is
  \begin{eqnarray}
  \bar{C}_{EE,\ell}^{(\rm S)}\Big|_{\rm DM}&=& (4\pi)^{2} \frac{(\ell+2)!}{(\ell-2)!} \int d^{3}K P_{S}(K) \left[ \dfrac{3}{4}\int_{0}^{\eta_{0}}d\eta \, g_{e}(\eta)\, \Pi(K,\eta)\, \dfrac{j_{\ell}(x)}{x^{2}} \right]^2 ,
  \end{eqnarray}
  \color{black}
  Eqs. \eqref{ClEE}, \eqref{ClBB}and \eqref{ClEB} can be approximated as follows:
  \begin{eqnarray}
  C_{EE,\ell}^{(\rm S)}\Big|_{\rm DM}&\cong& \cos^{2}(\tau_{DM}) \bar{C}_{EE,\ell}^{(\rm S)},\nonumber\\
  C_{BB,\ell}^{(\rm S)}\Big|_{\rm DM}&\cong& \sin^{2}(\tau_{DM}) \bar{C}_{EE,\ell}^{(\rm S)},\nonumber\\
  C_{EB,\ell}^{(\rm S)}\Big|_{\rm DM}&\cong& \dfrac{1}{4}\sin^{2}(2\tau_{DM}) \bar{C}_{EB,\ell}^{(\rm S)}.
  \end{eqnarray}
     The term $ \tau_{DM}$ is the effective opacity of DM produced by the DM-photon interaction from the last scattering surface to the present time and is determined by \eqref{DDM3} and \eqref{BE6}. 
 It is more convenient to express Eq.\eqref{BE6} in terms of the redshift $z$ which will be as follows:
   \begin{equation}\label{BE9}
   \tau_{\text{\tiny{DM}}}(z)=\,\frac{{\mathcal{M}}^2}{m_{\text{\tiny{DM}}}}\,\rho _{\text{\tiny{DM}}}^{0}\,\int_{0}^{z}dz^{'}\,\dfrac{(1+z^{'})^{2}}{H(z^{'})}.
   \end{equation}
    To arrive the above equation, we used
   \begin{equation}
   \rho _{\text{\tiny{DM}}}=\rho _{\text{\tiny{DM}}}^0 (1+z)^3,
   \end{equation}
   where $\rho _{\text{\tiny{DM}}}^0$ is the mass density of DM in the present time and
   \begin{equation}\label{redshift}
   	a\,d\eta=-\frac{dz}{H(z)(1+z)},
   \end{equation}
      where $H(z)$ is the Hubble parameter. Making use of the Friedmann equation in the matter dominated era  
   \begin{equation}\label{Hubble}
   \frac{H^2}{H_{0}^2}=\Omega_{M}^0(1+z)^3+\Omega_{\Lambda}^0,
   \end{equation}
     \begin{figure}[t!]
 	\includegraphics[scale=0.6]{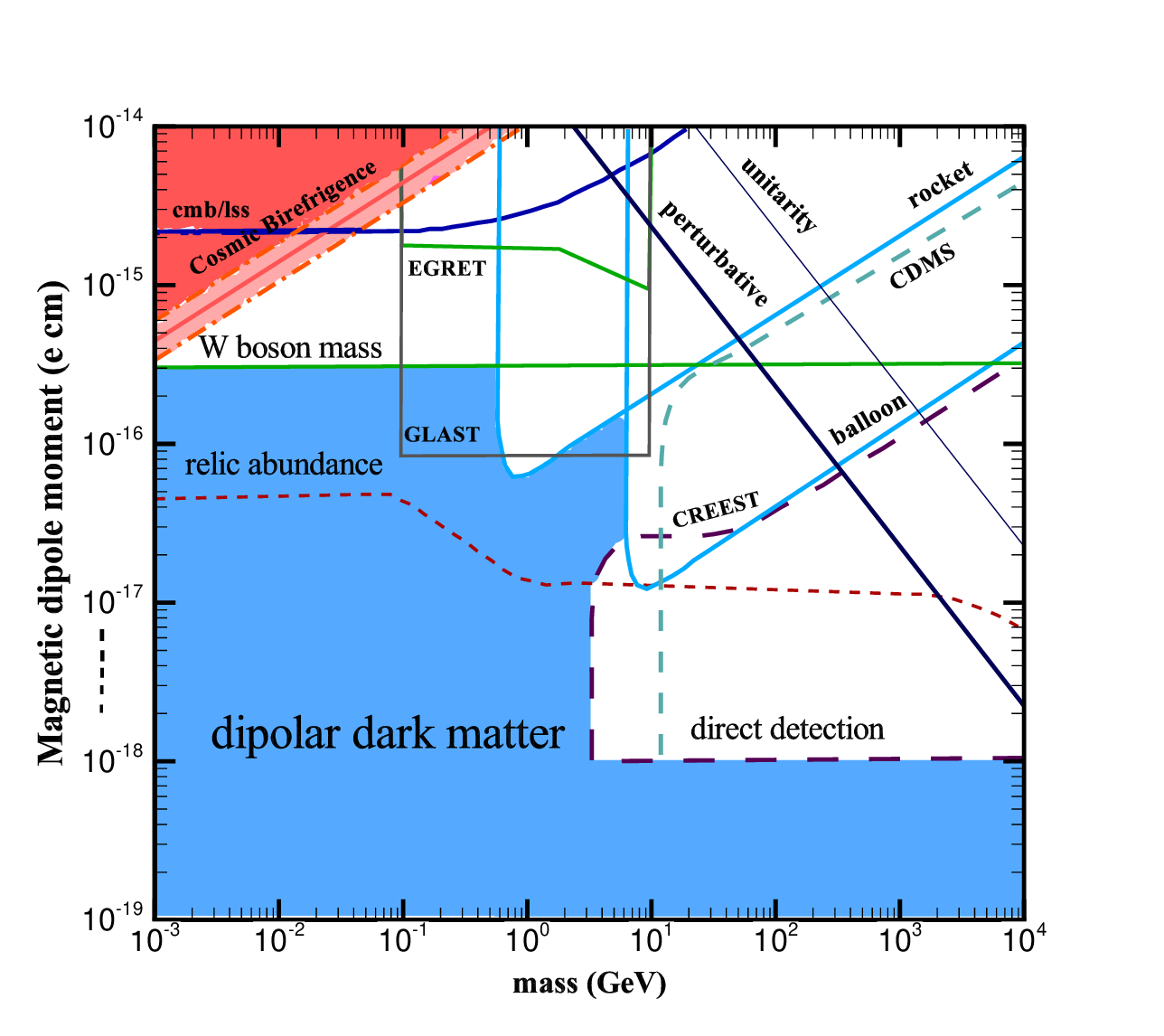}\\
 	\caption{(color online). Dipolar DM parameter space $[m,\mathcal{M}]$. The shaded blue region is the place where the viable Dirac candidates must lie in, below the solid lines and outside the long-dashed lines.
 		The short-dashed relic abundance curve, which is obtained by considering the standard freeze-out, indicates where the DM would meet a cosmological density $\Omega h^2=0.135$,
 assuming no particle-antiparticle asymmetry, and no interactions with standard-model particles except the dipole coupling to photons. Note that the EGRET and GLAST curves constrain the combination $(\mathcal{D}^4+\mathcal{M}^4)^{1/4}$; the perturbative and unitarity curves apply to the stronger of $(\mathcal{D}, \mathcal{M})$; while all other curves restrict $(\mathcal{D}^2+\mathcal{M}^2)^{1/2}$. (see \cite{Sigurdson:2004zp} for more detail). The plot of cosmic birefringence constrains $\mathcal{M}$. The shaded pink region is the area excluded from the dipolar DM parameter space by the results of CB for the Majorana dipolar DM. Moreover, the pale pink area marked by the dashed-dotted line represents the uncertainty in the upper limit obtained due to this effect.}\label{fig2}
 \end{figure}
The effective opacity \eqref{BE9} becomes
   \begin{equation}
   	\tau_{DM}=\frac{{\mathcal{M}}^2}{m_{\text{\tiny{DM}}}}\,\rho _{\text{\tiny{DM}}}^{0}\frac{2H(z')}{3\Omega_M^0H_0^{2}}\Big|^{z'=z}_{z'=0},
   \end{equation}
   \color{black}
   where $H_{0}\,=\, (67.4\,\pm\,0.5)\, \text{km}\text{s}^{-1}\text{Mpc}^{-1}$,\,\,$\Omega_{M}^0\,=\,0.315\,\pm\,0.007$, \,\,$\Omega_{\Lambda}^0\approx0.69$ \cite{Planck:2018vyg}.
 The redshift dependence of the above equation indicates that the maximum value of $\tau_{DM}$ occurs near the last scattering surface and its approximate value is as follows:
  \begin{equation}\label{tauDM}
  	\tilde{\tau}_{DM}\approx 4.6\times10^{25}\,\left(\frac{\mathcal{M}}{e\,\text{cm}}\right)^{2}\,\left(\frac{GeV}{m _{\text{\tiny{DM}}}}\right)\,\,\,\,\,\,\,\,\,
  	,\bigg(\frac{\rho _{\text{\tiny{DM}}}^{0}}{2.5\times 10^{-30\,\,g/cm^{3}}}\bigg)\,\,\bigg(\frac{z'}{10^3}\bigg).
  \end{equation}
  Now, using the equation between the CB angle $\beta$ and the maximum value of the effective opacity \cite{Mohammadi:2021xoh}
     \begin{equation}\label{bet}
  	\beta\,\approx\,\frac{1}{2}\,\tilde{\tau}_{DM},
  \end{equation}
we can estimate the contribution of the dipolar DM interaction with the photon in producing the CB effect. For instance, the CB angle of the CMB due to the interaction with dipolar DM whose mass is around $1 \text{MeV}$ will be approximated as   
\begin{equation}\label{beta}
\beta \big|_{\text\tiny{DM}}\approx 2.0\times 10^{-3}\,rad\,\,\,\,\,\left(\frac{\mathcal{M}}{3\times 10^{-16}\,e\,cm}\right)^2\,\,\,\left(\frac{10^{-3}GeV}{m_{\text{\tiny{DM}}}}\right).
\end{equation} 
Considering the value of the CB angle reported by using the Planck data, $\beta=0.30^\circ\pm0.11^\circ$ ($68\%\text{C.L.}$) \cite{Diego-Palazuelos:2022dsq}, we realize that dipolar DM with the mentioned properties can approximately compensate for about $(40\pm 13)\%$ of the CB angle of the CMB, where the uncertainty originates from the uncertainty on $\beta$. It is important to note that the chosen values for mass and magnetic moment in Eq. \eqref{beta} are threshold values, which means that according to cosmological constraints, the mass of dipolar DM must be larger than $1 MeV$ \cite{Burles:1999zt}, and also based on the reported results concerning the magnetic moment of DM, the value of this quantity is approximately less than $10^{-15} e cm$, (see Fig. \ref{fig2}) . Therefore, using relations Eqs. \eqref{tauDM} and \eqref{bet}, one can easily find that with the increase in mass and decrease in coupling constant, the contribution of this type of DM on the generation of the CB effect will be less, and hence, it can contribute to producing CB up to the mentioned value at most.\par
Another perspective that we can follow concerning this issue is putting a constraint on the magnetic dipole moment depending on the mass of DM particles. In this regard,
by combining Eqs. \eqref{tauDM} and \eqref{bet},
\begin{equation}
	\mathcal{M}\approx \big(\frac{1}{2.3\times 10^{25}}\big)^{1/2}\,\sqrt{\beta}\,\,\,e\text{\,cm}\,\,\,\,\,\,\,\,\,\,\,\,\big(\frac{m _{\text{\tiny{DM}}}}{1\,GeV}\big)^{1/2},
\end{equation}
\color{black}
  and substituting the value of CB angle reported by using the Planck data release, $\beta=0.30^\circ\pm0.11^\circ$ ($68\%\text{C.L.}$), the following constraint will be placed on the phase parameters of the dipolar DM 
  \begin{equation}
  	\mathcal{M}\leqslant(1.4\pm 0.23)\times10^{-14}\, e\text{\,cm}\,\,\,\,\,\,\,\,\,\,\,\,\,\,\,\big(\frac{m _{\text{\tiny{DM}}}}{1\,GeV}\big)^{1/2},
  \end{equation}
    where the uncertainty on $\mathcal{M}$ originates from the uncertainty on $\beta$, i.e.
     \begin{equation}
     \Delta\mathcal{M}\approx\big(\frac{\beta}{2.3\times 10^{25}}\big)^{1/2}\,\,\frac{\Delta\beta}{2\beta}\,\,\,\,\,\,\,\,e\text{\,cm}\,\,\,\,\big(\frac{m _{\text{\tiny{DM}}}}{1\,GeV}\big)^{1/2}
     \end{equation}
    
  \color{black}
  For instance, our results put a bound on the magnetic dipole moment about $\mathcal{M}\leqslant (7\pm 1.1)\times10^{-16}\,e\text{\,cm}$ for dipolar DM particles whose mass is around $m _{\text{\tiny{DM}}} \approx 3\,MeV$.
 The full mass dependence of this result is shown in Fig.\ref{fig2}. This figure is originally based on the constraints on the dipolar DM parameter space that come from some theoretical and experimental research and are adapted from \cite{Sigurdson:2004zp}. The Viable Dirac dipolar DM must lie in the shaded blue region, below the solid lines, and outside the long-dashed lines (see \cite{Sigurdson:2004zp} for more detail).
 The shaded pink region is the region excluded from the dipolar DM parameter space by the CB results. It is important to mention that the result obtained from considering the CB effect is related to Majorana dipolar DM particles, while other results are dedicated to Dirac dipolar DM. Indeed, we put a constraint on the sub-GeV Majorana dipolar DM through the CB effect of the CMB photons. It is also notable that the constraints obtained on Majorana dipolar DM, including the constraints that come from the direct detection experiments \cite{Masso:2009mu}, study DM particles whose mass is around GeV or higher, while the constraint that we obtain here is regarding the sub-GeV DM particles. Before ending this section, it is worth mentioning that although an experiment like W boson mass excludes the sub-GeV Dirac dipolar DM with a magnetic moment larger than $7\times10^{-16}\,e\,\text{cm}$, part of this region will be accessible for Majorana dipolar DM based on the CB effect of the CMB.
    \color{black}
  \subsection{Sterile Neutrino dark matter}
    The existence of the right-handed sterile neutrinos is elegantly formulated in the seesaw model. In the framework of type-I seesaw model, the SM is extended by at least two heavy sterile neutrino singlets $\nu_\text{\tiny{R}}^i$ ($i$ indicates the generation) which mix with the SM neutrinos through a mixing angle $\theta$ and form the neutrino physical states as follows:
    \begin{eqnarray}\label{masseig}
   	N =V^\dagger_\text{\tiny{N}}\nu_\text{\tiny{R}}+U^{\dagger}_{N}\theta\nu^c_\text{\tiny{L}}+h.c.\,, ~~~~\text{and}~~~~
    	\upnu = V^\dagger_\nu \nu_\text{\tiny{L}}-U^\dagger_\nu \theta \nu_\text{\tiny{R}}^c+h.c.\,,
    \end{eqnarray} 
    where $V$, $\theta$ and $U$ contain information about the mixing angle \cite{Haghighat:2019rht}. Indeed, $V_\nu$ is the usual neutrino mixing matrix connecting the observed light mass eigenstates to the active flavor eigenstates:
    \begin{equation}
    V_\nu\equiv (1-\frac{1}{2}\theta\theta^\dagger)U_\nu,
    \end{equation}
    and $U_\nu$ is the unitary part of neutrino mixing matrix \cite{Boyarsky:2018tvu}. Meanwhile, the corresponding parameters in the sterile sector are $V_\text{\tiny{N}}$ and $U_\text{\tiny{N}}$ and the active-sterile mixing  angle is 
    \begin{equation}
    \Theta\equiv \theta U_\text{\tiny{N}}^\star.
    \end{equation} 
       In the SM, neutrinos interact with other particles only via the weak interaction,
    \begin{equation}\label{WeakWW}
    -\frac{g}{\sqrt{2}}\overline{{\nu_l}_{\tiny{L}}}\gamma^\mu l_L W^+_\mu
    -\frac{g}{\sqrt{2}}\overline{l_L}\gamma^\mu {\nu_l}_L W^-_\mu  
    - \frac{g}{2\cos\theta_W}\overline{{\nu_{l}}_L}\gamma^\mu{\nu_{l}}_L Z_\mu ,
    \end{equation}
     where $g$ is the gauge coupling constant, $\theta_{W}$ stands for the Weinberg angle, $l=e, \mu, \tau$ and, ${\nu_l}_{L}$ denotes the flavor state of left-handed SM neutrinos. Using Eqs.\eqref{masseig}, one can express the neutrino flavor eigenstates ($\nu_{L}$) in terms of neutrino physical states, i.e. $\nu_L=V_\nu \upnu + \Theta N$ and 
    therefore, in the most general form, the mass eigenstate sterile neutrinos($N$) can interact with the SM particles through the mixing angle as follows \cite{Boyarsky:2018tvu}:
    \color{black}
     \begin{eqnarray}\label{SN}
   	\mathcal{L}&\supset&\,\,\sum_{l}-\frac{g}{\sqrt{2}}\bar{N}\,\Theta^{\dagger}\,\gamma ^{\mu}\,l_{\text{\tiny{L}}}\,W_{\mu}^{+}\,-\sum_{l}\frac{g}{\sqrt{2}}\,\bar{l}_{\text{\tiny{L}}}\,\gamma^{\mu}\,\Theta\,N\,W_{\mu}^{-}-\frac{g}{2\cos \theta_{\text{\tiny{W}}}}\,\bar{N}\,\Theta^{\dagger}\,\gamma ^{\mu}\,\nu_{l_{\text{\tiny{L}}}}\,Z_{\mu}\,\nonumber\\
   	&&-\frac{g}{2\cos \theta_{\text{\tiny{W}}}}\,\bar{\nu}_{l_{\text{\tiny{L}}}}\,\gamma^{\mu}\,\Theta\,N\,Z_{\mu}-\frac{g}{\sqrt{2}}\,\frac{M_{\text{\tiny{N}}}}{m_{\text{\tiny{W}}}}\,\Theta\,h\,\bar{\nu}_{l_{\text{\tiny{L}}}}\,N\,-\frac{g}{\sqrt{2}}\,\frac{M_{\text{\tiny{N}}}}{m_{\text{\tiny{W}}}}\,\Theta^{\dagger}\,h\,\bar{N}\,\nu_{l_{\text{\tiny{L}}}}.
    \end{eqnarray} 
    where $h$ is physical Higgs field, $M_{N}$ denotes the mass of sterile neutrino and $m_{W}$ stands for the mass of $W$ boson. Note that depending on models and the sterile neutrino production mechanism, and considering the astrophysical constraints, one can find different bounds on the mixing angle.
     For instance, some considerations which predict active generation
    of such particles in the early Universe constrain $\theta^2\ll 10^{-8}$ from the total DM relic density and the absence of X-ray signal from sterile neutrino decay\cite{{Boyarsky:2018tvu},{Dolgov:2002wy}}. However, regarding some models with a hidden sector coupled to the
    sterile neutrino, these bounds can be extended to $\theta^2\leqslant 10^{-1}$ from the total DM relic density \cite{Bezrukov:2017ike}. \par
 
\begin{figure}[tb]
 		\center
 		\includegraphics[scale=1]{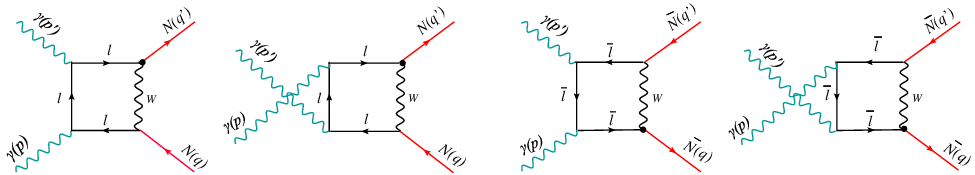}\\
 		\caption{The representative Feynman diagrams represent the photon-Sterile neutrino scattering, where $l=e, \mu,\tau$ and $\bar{l}$ indicates the anti-particle of $l$.} \label{feyn}
 	\end{figure}

 Based on some reports regarding the galaxy phase space density, universal galaxy surface density and the DM density, the mass eigenstate sterile neutrinos (N) can be fit to a warm DM scenario \cite{Dodelson:1993je}.
 Here, we study the CB of the CMB due to its interaction with sterile neutrino DM.  In the context of the seesaw model, photons can scatter from sterile neutrinos at a one-loop level with a lepton (or antilepton) and weak gauge bosons propagating in the loop. As the authors have shown in \cite{Haghighat:2019rht}, the sterile neutrino-CMB interaction given by Fig.\ref{feyn} can affect the CMB polarization features and modify the power spectrum of the B- mode polarization. Indeed, to examine the effects of the photon-sterile neutrino interaction on the polarization of the CMB photons, we take (\ref{SN}) and (\ref{forward}) into account to find the time evolution of the density matrix components as follows (see appendix B for more detail) \cite{Haghighat:2019rht}:
     \begin{eqnarray}\label{rhodot}
     \frac{d}{dt}\rho_{ij}(k)\,&=&\,-\frac{\sqrt{2}}{12\,\pi\,k^{0}}\alpha\,\theta^{2}\,{G}_{\text{\tiny{F}}}\int d{\bf{q}}\,\, (\delta_{is}\rho_{s'j}(k)-\delta_{js'}\rho_{is}(k))\,f_\text{\tiny{DM}}({\bf{x}},{\bf{q}})\,
     \bar{u}_{r}(q)\,\,(1-\gamma^{5})\nonumber\\&&(q\cdot\epsilon_{s}\,\,\slashed{\epsilon}_{s^{'}}\,+\,q\cdot\epsilon_{s^{'}}\,\,\slashed{\epsilon}_{s})\,u_{r}(q)+\frac{\sqrt{2}}{24\,\pi\,k^{0}}\alpha\,\theta^{2}\,{G}_{\text{\tiny{F}}}\int d{\bf{q}}\,\, (\delta_{is}\rho_{s'j}(k)-\delta_{js'}\rho_{is}(k))\,\nonumber\\&&f_\text{\tiny{DM}}({\bf{x}},{\bf{q}})\,    \bar{u}_{r}(q)\,(1-\gamma^{5})\,\slashed{k}\,(\slashed{\epsilon}_{s^{'}}\,\slashed{\epsilon}_{s}\,-\,\slashed{\epsilon}_{s}\,\slashed{\epsilon_{s^{'}}})\,u_{r}(q).
     \end{eqnarray}
    where $\epsilon_{s}(k)$ with $s=1,2$ are the photon polarization 4-vectors of two physical transverse polarizations, while $u_{r}(q)$ and $v_{r}(q)$ are the Dirac spinors. Furthermore, $f_{\text{\tiny{DM}}}$ denotes the distribution function of DM, and $G_{F}$ and $\alpha$ are the Fermi coupling constant and electromagnetic fine structure constant, respectively. One can reconstruct the Stokes parameters through the density matrix elements
    and using the following identities
    \begin{eqnarray}
   	\sigma_{\mu\nu}\gamma_{\alpha}&=&-i(\delta_{\mu\alpha}\gamma_{\nu}+\delta_{\nu\alpha}\gamma_{\mu}+\epsilon_{\mu\nu\alpha\lambda}\gamma^{\lambda}\gamma^{5}),\nonumber\\
    	\bar{u}_{r}(q)\gamma^{\mu}\,u_{r}(q)&=& 2\frac{q^{\mu}}{m_{DM}},\nonumber\\
    	\bar{u}_{r}(q)\gamma^{\mu}(1\pm \gamma^5)\,u_{r}(q)&=& 2\frac{q^{\mu}}{m_{DM}},
    \end{eqnarray}
    where the completely antisymmetric alternating symbol $\epsilon_{\mu\nu\alpha\lambda}$ is equal to $+1$ for ($\mu,\nu,\alpha,\lambda$) 
    an even permutation of $(0, 1, 2, 3)$, is equal to $-1$ for an odd permutation, and vanishes if two or more indices are the same. 
    Consequently, reconstruction of the Stokes parameters shows that this interaction can affect the evolution of the linear polarization of the CMB as follows:
     \begin{equation}\label{B mode}
     \frac{d}{d\eta}\,\Delta_{P}^{\pm(S)}\,+\,i\,K\,\mu\,\Delta_{P}^{\pm(S)}\,=\,C_{e\gamma}^{\pm}\,\mp\,i\,a(\eta)\dot{\tau}_\text{\tiny{DM}}\,\Delta_{P}^{\pm},
     \end{equation}
      in which $\dot{\tau}_\text{\tiny{DM}}$ considered for the contribution of the photon-sterile neutrino scattering can be obtained as
     \begin{eqnarray}\label{eta}
     \dot{\tau}_\text{\tiny{DM}}&=&\frac{\sqrt{2}}{3\pi k^{0}\,m_{\text{\tiny{DM}}}}\,\,\alpha\,\,{G}_{\text{\tiny{F}}}\,\theta^{2}\,\int\,d{\bf{q}}\,f_\text{\tiny{DM}}({\bf x},{\bf q})\,\times (\varepsilon _{\mu\,\nu\,\rho\,\sigma}\epsilon_{2}^{\mu}\,\epsilon_{1}^{\nu}\,k^{\rho}\,q^{\sigma}),
     \end{eqnarray}
    where it can be reduced to 
    \begin{eqnarray}
    \dot{\tau}_\text{\tiny{DM}}&=&  \frac{\sqrt{2}}{3\pi k^{0}\,m_{\text{\tiny{DM}}}}\,\,\alpha\,\,{G}_{\text{\tiny{F}}}\,\theta^{2}\int d\mathbf{q}\, f_\text{\tiny{DM}}({\bf x},{\bf q})
    \times \left[q^0 {\bf k}\cdot(\epsilon_1\times\epsilon_2)+k^0 {\bf q}\cdot(\epsilon_1\times\epsilon_2)\right]\nonumber\\
    &=&  \frac{\sqrt{2}}{3\pi}\,\,\alpha\,\,{G}_{\text{\tiny{F}}}\,\theta^{2}\,n_{\text{\tiny{DM}}}\left[1+\langle {\bf v}\rangle\cdot(\epsilon_1\times\epsilon_2)\right]\approx \frac{\sqrt{2}}{3\pi}\,\,\alpha\,\,{G}_{\text{\tiny{F}}}\,\theta^{2}\,n_{\text{\tiny{DM}}},\label{Bmode1}
    \end{eqnarray}
    where ${\bf k}\cdot(\epsilon_1\times\epsilon_2)= |{\bf k}|$, the DM number density $n_\text{\tiny{DM}}=\int \frac{d^3{\bf{q}}}{(2\pi)^3} f_\text{\tiny{DM}}({\bf x},{\bf q})$, and $ \langle {\bf v}\rangle $ is the average velocity of DM particles. Since the average velocity of DM is small, the dominated contribution of
this scattering to photon polarization comes
from the first term and thus we ignore the term including $ \langle {\bf v}\rangle $.
 \color{black}
 
Using \eqref{BE6}, \eqref{redshift} and \eqref{Hubble}, we arrive at the following equation for the effective opacity $\tau_{\text{\tiny{DM}}}$
 \begin{equation}
 \tau_{\text{\tiny{DM}}}=\frac{\sqrt{2}}{3\pi\,m_{\text{\tiny{DM}}}}\,\,\alpha\,\,{G}_{\text{\tiny{F}}}\,\theta^{2}\,\rho_{\text{\tiny{DM}}}^{0}\,\frac{2H(z')}{3\Omega_M^0H_0^{2}}\Big|^{z'=z}_{z'=0},
 \end{equation}
 \color{black}
where we have used the fact that $\rho_{\text{\tiny{DM}}}\,=\,\rho _{\text{\tiny{DM}}}^0 (1+z)^3$ in which $\rho _{\text{\tiny{DM}}}^0$ is the mass density of DM in the present time. Now, we try to make an estimate of the maximum value $\tau_{\text{\tiny{DM}}}$ near the last scattering which leads to the following equation:
\begin{equation}
\tilde{\tau}_{\text{\tiny{DM}}}\approx 3\times 10^{-9}\,\theta^{2}\,\left(\frac{GeV}{m_{\text{\tiny{DM}}}}\right)\,
\bigg(\frac{\rho _{\text{\tiny{DM}}}^{0}}{10^{-47}\,GeV^{4}}\bigg)\,\,\bigg(\frac{z'}{10^3}\bigg).
\end{equation} 
Hence, the CB angle of the CMB due to the interaction with the sterile neutrinos can be approximated as 
\begin{equation}
\beta \approx 1.5\times 10^{-9}\,\theta^{2}\,\left(\frac{GeV}{m_{\text{\tiny{DM}}}}\right).
\end{equation} 

Before to proceed, it is worth discussing the approximate CB angle which can be caused by this sort of interaction. Since according to some cosmological constraints, the mass of sterile neutrinos must be larger than $100 eV$ \cite{Boyarsky:2018tvu} and based on the reported results regarding the mixing angle, the value of this quantity is approximately less than $10^{-3}$, we come to the conclusion that by considering the value of the CB angle, $\beta=0.30^\circ\pm0.11^\circ$, around $(0.30\pm 0.10)\%$ of the CB angle can be caused by the interaction with the sterile neutrino DM (if it exists).
\color{black}
 \begin{figure}
 	\includegraphics[scale=0.6]{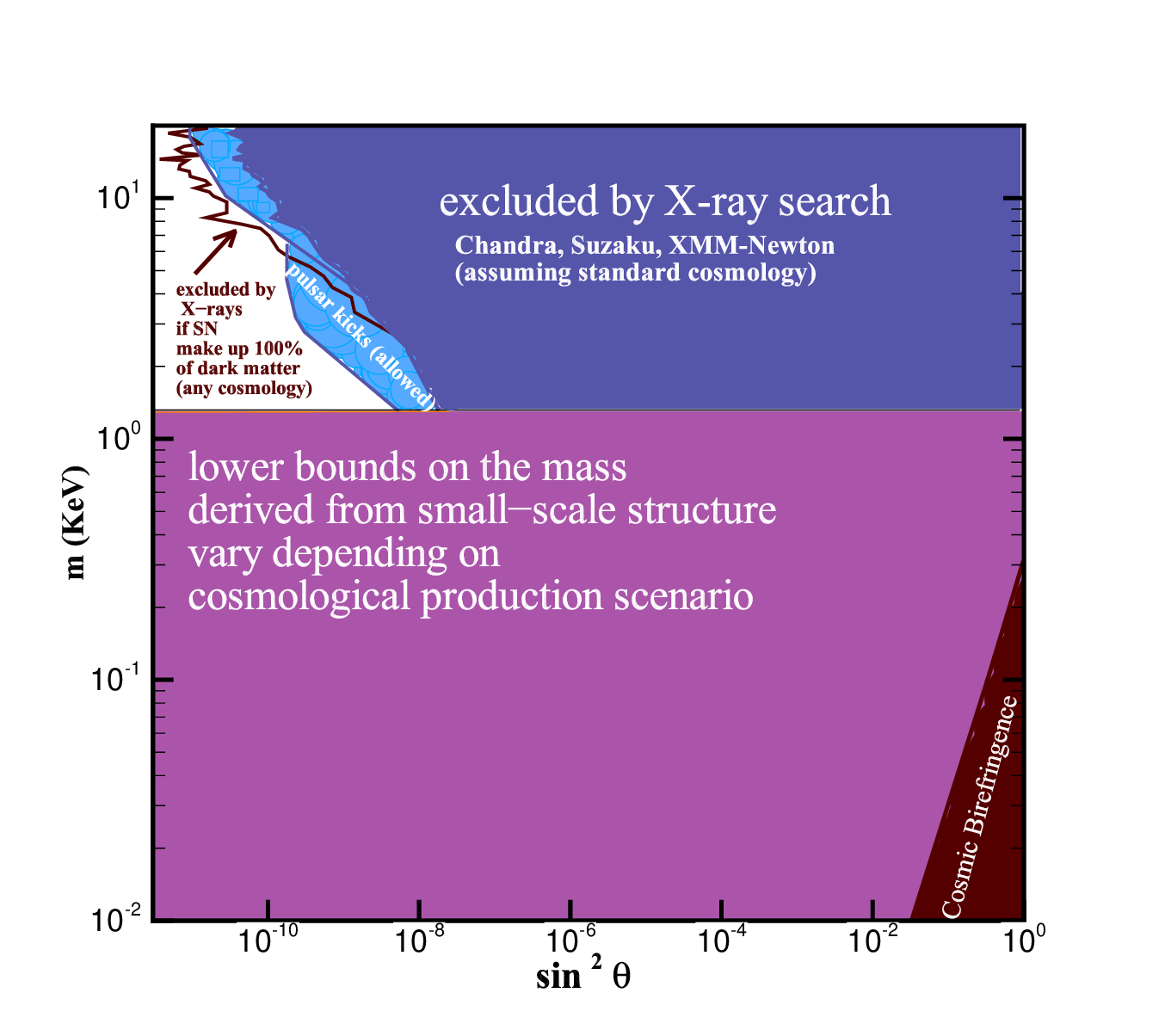}\\
 	\caption{(color online). The constraints on the sterile neutrino DM parameter space assuming a standard cosmology below the temperature when neutrino oscillations occur are adapted from \cite{Buckley:2013bha}. The shaded brown region is the area excluded by the CB effect of the CMB.}\label{fig3} 
 \end{figure}
Moreover, using the mentioned value reported for the CB by the Planck collaboration, one can put a constraint on the parameter space of the sterile neutrino DM:
\begin{equation}
	\theta^2\leq\,(3.3\pm 1.1)\,(rad)^2\,\,\,\,\,\,(\frac{m_{\text{\tiny{DM}}}}{1\,\text{KeV}})\,
\end{equation}
The parameter space of the sterile neutrino DM $[m,\sin ^{2}\theta]$ is depicted in Fig. \ref{fig3}. This figure, adapted from \cite{Buckley:2013bha}, is originally based on the summary of astrophysical constraints on the parameter space, $m_{st}-\theta$ plane, for the sterile neutrino DM. 
As this figure shows, the constraint on the sterile neutrino parameter space due to the CB effect is placed in the region that has already been excluded by the X-ray experiment. However, we should point out that the values of mass ($m_{\text{\tiny{DM}}}$) and the mixing angle ($\theta$) could depend on cosmological production scenarios. In fact, one of the most important issues of sterile neutrinos is to address the question of how they have been produced in the early universe. Generally, sterile neutrinos can be produced by neutrino oscillation in the primordial plasma via a tiny active-sterile neutrino mixing angle $\theta$, as first described by Dodelson and Widrow (DW) \cite{Dodelson:1993je}. Various current astrophysical observations impose severe constraints on the sterile neutrino DM which is generated by the DW mechanism \cite{Boyarsky:2005us,Boyarsky:2006fg,Boyarsky:2006ag,Boyarsky:2007ay,Yuksel:2007xh, Seljak:2006qw,Boyarsky:2008xj,Perez:2016tcq}. Meanwhile, more complex mechanisms for the production of DM, such as the Shi-Fuller mechanism \cite{Shi:1998km} which describes resonant oscillation production or other non-thermal production mechanisms including the decay of an extra-singlet scalar \cite{Shaposhnikov:2006xi,Kusenko:2006rh}, and scatterings through new mediators in the thermal bath without reaching thermal equilibrium \cite{Khalil:2008kp,Kaneta:2016vkq,Biswas:2016bfo,Seto:2020udg}, have been proposed. \par
 
 Before ending this section, we emphasize that if the production mechanism of the sterile neutrino DM is based on the DW mechanism, the area excluded by the results of the CB effect is placed in the region that was already ruled out by the X-ray experiment. However, if sterile neutrinos are produced through other mechanisms, masses less than 1 KeV and mixing angles larger than $10^{-4}$ are allowed; therefore, the constraint obtained from the results of the CB effect is a new constraint and excludes part of those areas.\par
 
 As a final point, it is worth mentioning that, here, we considered sterile neutrino DM in the context of seesaw model. However, it is possible to introduce the right-handed sterile
neutrinos as the DM candidates which can be coupled
effectively to the SM particles through the right-handed
current interactions with the SM intermediate gauge bosons \cite{{Xue:2016dpl},{Xue:2015wha},{Xue:1997tz},{Xue:1996fm}}. Indeed, this model
was motivated by the parity symmetry reconstruction
at high energies without any extra gauge bosons. This sort of DM candidates might also be considered as a new source of CB  which is under
investigation as a future work. 

 \color{black}
     \section{ Conclusion}\label{conclusion}
    We examined whether the existence of the CB angle of CMB photons could be used as a tool to study the nature and properties of DM. To this end, we considered two types of DM candidates, i.e. dipolar and sterile neutrino DM, and calculated the forward scattering contribution to investigate the CMB polarization effects on DM properties. We found that the interaction of those probable DM candidates and CMB photons results in generating the B mode polarization patterns and, consequently, produces the CMB CB effect. Using the birefringence angle reported by using the Planck data release, we discussed the properties of the mentioned candidates of DM, and the results are as follows: 
 \begin{itemize}
 	\item{Calculations performed regarding the dipolar DM showed that this DM candidate can contribute to generate a part of the CB effect of the CMB. Moreover, from another point of view, we used the reported CB angle of the CMB to put a new constraint on its electromagnetic coupling and mass, i.e., $\mathcal{M}/10^{-15}e.cm\approx 10\,\sqrt{m_{DM}/GeV}$, which means that for dipolar DM particles whose mass is about $m _{\text{\tiny{DM}}} \approx 3\,MeV$, the magnetic dipole moment will be around $\mathcal{M}\leqslant (7.6 \pm 1.1)\times10^{-16}\,e\text{\,cm}$. For further clarification, we provided Fig.\ref{fig2} in which the full mass dependence of the result has been illustrated. Note that this figure is originally adapted from ref\cite{Sigurdson:2004zp} and we added our result to it. Indeed, we put a constraint on the sub-GeV Majorana dipolar DM through the CB effect of the CMB photons. It is also notable that the constraints obtained on Majorana dipolar DM, including the constraints that come from the direct detection experiments \cite{Masso:2009mu}, study DM particles whose mass is around GeV and higher, while the constraint that we obtain here is concerning the sub-GeV DM particles.}
 		\item{In the case of the sterile neutrino DM, we found that this sort of DM candidate can also contribute to produce a part of the CB angle of the CMB. Furthermore, by using the reported CB angle of the CMB, the mixing angle $\theta^2$ and mass of the sterile neutrino are constrained as $\theta^2 \approx (3.3\pm 1.1)\,\,(rad)^2\, \frac{m_{DM}}{KeV}$. It seems that this constraint has already been excluded by some experiments such as X-ray experiment, which explains the exclusion based on the DW mechanism of sterile neutrino production. However, it should be pointed out that the values of mass ($m_{\text{\tiny{DM}}}$) and the mixing angle ($\theta$) could depend on the cosmological production scenarios. Besides the DW mechanism, more complex mechanisms for the production of DM, such as the Shi-Fuller mechanism or other non-thermal production mechanisms including the decay of an extra singlet scalar, and scatterings through new mediators in the thermal bath without reaching thermal equilibrium, have been proposed. It is notable that the importance of our result depends on the sterile neutrino production mechanism; if the production mechanism of the sterile neutrino DM is based on the DW mechanism, the area which is excluded by the results of the CB effect is placed in the region that already been ruled out by the X-ray experiment. Meanwhile, if sterile neutrinos are produced through other mechanisms, masses less than 1 KeV and mixing angles larger than $10^{-4}$ are allowed and, therefore, the constraint obtained from the results of the CB effect is a new constraint and excludes part of those areas.    			
     			
     			}
     		
     		      \end{itemize}
     
       \color{black}
       
   \section*{Acknowledgment}
   S. Mahmoudi is grateful to the Iran Science Elites Federation for
   the financial support.

   { \appendix
   \section{Studying the contribution of the
    	photon-dipolar DM scattering on the CMB polarization in case ${\Large{\delta\gg k^0}}$}
   In this appendix, we are going to calculate the contribution of the photon-Majorana dipolar DM forward scattering in case ${{\delta\gg k^0}}$.
      	In this case, Eq. (\ref{majorana122}) can be estimated as follows
   	\begin{eqnarray}\label{majorana16}
   	{M}\simeq \,\mathcal{M}_{12}^{2}\,\,\frac{(2k\cdot q)^{2}}{(m_{\text{\tiny{DM}}_{2}}^{2}-m_{\text{\tiny{DM}}_{1}}^{2})^{2}}\,\,{\bf{k}}\cdot(\vec{\epsilon}_{s'}\times\vec{\epsilon}_{s})+\,(1\,\leftrightarrow\,2).
   	\end{eqnarray}
   	Working in the non-relativistic limit (${\bf{q}}\approx m_{\text{\tiny{DM}}}$) and assuming $m_{\text{\tiny{DM}}_{1}}$ to be the same order of $m_{\text{\tiny{DM}}_{2}}$, one will arrive at the following relation
   	\begin{eqnarray}\label{majorana17}
   	{M}\simeq \,\mathcal{M}_{12}^{2}\,\,(\frac{k^{0}}{\delta})^{2}\,\,{\bf{k}}\cdot(\vec{\epsilon}_{s'}\times\vec{\epsilon}_{s})\,+\,(1\,\leftrightarrow\,2).
   	\end{eqnarray}
   	For the cases in which $k^{0}\ll\delta\ll m_{\text{\tiny{DM}}} $, $(m_{\text{\tiny{DM}}_{1}}\approx m_{\text{\tiny{DM}}_{2}}\approx m_{\text{\tiny{DM}}})$, and after some calculation, one can find the evolution of  the Stokes parameters similar to \eqref{inten}-\eqref{circpol} except that $\dot{\tau}_{\text{\tiny{DM}}}$ is defined as follows
   	 \begin{equation}
   	 \dot{\tau}_{\text{\tiny{DM}}}=\,(\frac{k^{0}}{\delta})^{2}\frac{ \mathcal{M}^2}{m_{\text{\tiny{DM}}}}\, \rho_{\text{\tiny{DM}}}.
   	 \end{equation}
   	   	The above relation clearly shows that the contribution of the
   	   	photon-dipolar DM scattering on the CMB polarization will be suppressed as $(\frac{k^0}{\delta})^2$ compared to the case $\delta\ll k^0$ and therefore, we will not consider this case in this paper. \par

\section{Calculation of Time evolution of the density matrix components via photon-Sterile neutrino interaction}
This appendix aims to calculate the time evolution of the density matrix components due to the forward scattering of the photon-Sterile neutrino interaction. To this end, we use the seesaw Lagrangian given in (\ref{SN}) to find the types of possible interactions between sterile neutrinos and photons. Indeed, the dominant interaction comes from the scattering of photons from Sterile neutrinos at a one-loop level with a lepton and weak gauge bosons propagating in the loop. Representative relevant Feynman diagrams are shown in Fig. \ref{feyn}.\par

Fourier transformations of the electromagnetic free gauge field $A^{\mu}$ and Majorana fermion field $N(x)$, which are self-conjugate, are as follows
\begin{equation}\label{gauge}
A_{\mu}(x)=\int\frac{d^3 {\bf k}}{(2\pi)^3 2k^0}[a_{s}(p)\epsilon_{s\mu}(k)e^{-ik.x} + a_{s}^\dagger (k)\epsilon_{s\mu}^* (k)e^{ik.x}],
\end{equation}
\begin{equation}\label{majorana1}
N(x)= \int \frac{d^3{\bf q} }{(2\pi)^3}\frac{m_{\tiny{\text{DM}}}}{ q^0}\left[ b_r(q) u_{r}(q)
e^{-iq\cdot x}+ b_r^\dagger (q) v_{r}(q)e^{iq\cdot x}
\right],
\end{equation}
where $\epsilon_{s\mu}(p)$  with $s=1,2$ are the photon polarization 4-vectors of two physical transverse polarization while $u_{r}(q)$ and $v_{r}(q)$ are the Dirac spinors. The creation $a_{s}^\dagger(k)$ ($b_{r}^\dagger(q)$) and annihilation $a_{s}(k)$ ($b_{r}(q)$) operators respect the following canonical commutation (anti-commutation) relations
\begin{eqnarray}
[a_{s}(k),a_{s'}^\dagger (k')]&=&(2\pi)^3 2k^0 \delta_{ss'}\delta^{(3)}({\bf k}-{\bf k'}),\nonumber \\
\{b_{r}(q),b_{r'}^\dagger (q')\} &=&(2\pi)^3 \frac{q^0}{m_{\tiny{\text{DM}}}} \delta_{rr'} \delta^{(3)}({\bf q}-{\bf q'}).
\end{eqnarray}
Making use of Eqs. \eqref{HI}, \eqref{SN}, and the above relations, one can find that the leading-order interacting Hamiltonian for the scattering representing in Fig. \ref{feyn} can be expressed by the following scattering amplitude
\begin{eqnarray}\label{H1}
H_{I}^0(t)&=&\,\int {d\bf{q}} {d\bf{q'}} {d\bf{k}} {d\bf{k'}}(2\pi)^3 \delta^{(3)} ({\bf{q'}}+{\bf{k'}}-{\bf{q}}-{\bf{k}}) \exp({i[q'^0 +k'^0 -q^0 - k^0]})\nonumber\\
&\times&[b_{r'}^\dagger({\bf q'}) a_{s'}^\dagger({\bf k'})\mathcal{M}_\text{tot}(N\gamma\,\to\,N\gamma)\, a_{s}({\bf k}) b_{r}({\bf q})],
\end{eqnarray}
with ${d \bf{q}} \equiv \frac {d^3 \bf{q}}{(2\pi)^3}\frac{\tiny{\text{DM}}}{q^0}$, ${d \bf{k}} \equiv \frac {d^3 \bf{k}}{(2\pi)^3}\frac{1}{2 k^0}$ and the total amplitude ${M}_\text{tot}$ can be obtained from the sum of all Feynman diagrams in Fig.\ref{feyn}, as follows 
\begin{eqnarray}\label{m}
{M}_{tot}({\bf q'}r',{\bf k'}s',{\bf q}r,{\bf k}s)&\equiv &{M}_{1}({\bf q'}r',{\bf k'}s',{\bf q}r,{\bf k}s) +{M}_{2}({\bf q'}r',{\bf k'}s',{\bf q}r,{\bf k}s)\nonumber\\
&&-{M}_{3}({\bf q'}r',{\bf k'}s',{\bf q}r,{\bf k}s) -{M}_{4}({\bf q'}r',{\bf k'}s',{\bf q}r,{\bf k}s),
\end{eqnarray}
where ${M}_{3,4}({\bf q'}r',{\bf k'}s',{\bf q}r,{\bf k}s)$ are, respectively,  the Hermitian conjugates of ${M}_{1,2}({\bf q'}r',{\bf k'}s',{\bf q}r,{\bf k}s)$ and have been contributed from  antiparticles in the loops as follows
\begin{eqnarray}\label{m1}
{M}_{1}({\bf q'}r',{\bf k'}s',{\bf q}r,{\bf k}s)&=&\frac{1}{(2\pi)^{4}}\,\frac{e^{2}\,g^{2}}{8}\,\theta^{2}\int d^{4}l\,\, \,\bar{u}_{r^{'}}({\bf q^{'}})\gamma^{\alpha}\,(1-\gamma^{5})\,S_{F}(l+k-k^{'})\slashed{\epsilon}_{s^{'}}({\bf k^{'}})\nonumber\\
&&S_{F}(k+l)\,\slashed{\epsilon}_{s}({\bf k})\,S_{F}(l)\gamma ^{\beta}\,(1-\gamma^{5})\,u_{r}({\bf q})\,D_{F_{\alpha\beta}}(q-l),
\end{eqnarray}
\begin{eqnarray}\label{m2}
{M}_{2}({\bf q'}r',{\bf k'}s',{\bf q}r,{\bf k}s)&=&\frac{1}{(2\pi)^{4}}\,\frac{e^{2}\,g^{2}}{8}\,\theta^{2}\int d^{4}l\,\, \,\bar{u}_{r^{'}}({\bf q^{'}})\gamma^{\alpha}\,(1-\gamma^{5})\,S_{F}(l+k-k^{'})\slashed{\epsilon}_{s}({\bf k})\nonumber\\&&S_{F}(l-k^{'})\,\slashed{\epsilon}_{s^{'}}({\bf k^{'}})\,S_{F}(l)
\gamma ^{\beta}\,(1-\gamma^{5})\,u_{r}({\bf q})\,D_{F_{\alpha\beta}}(q-l),
\end{eqnarray}
\begin{eqnarray}\label{m3}
{M}_{3}({\bf q'}r',{\bf k'}s',{\bf q}r,{\bf k}s)&=&\frac{1}{(2\pi)^{4}}\,\frac{e^{2}\,g^{2}}{8}\,\theta^{2}\int d^{4}l\,\, \,\bar{v}_{r}({\bf q})\gamma^{\alpha}\,(1+\gamma^{5})\,S_{F}(-l)\slashed{\epsilon}_{s}({\bf k})S_{F}(-k-l)\nonumber\\
&&\,\slashed{\epsilon}_{s^{'}}({\bf k^{'}})\,S_{F}(k^{'}-k-l)\gamma ^{\beta}\,(1+\gamma^{5})\,v_{r}({\bf q'})\,D_{F_{\alpha\beta}}(l-q),
\end{eqnarray}
and 
\begin{eqnarray}\label{m4}
{M}_{4}({\bf q'}r',{\bf k'}s',{\bf q}r,{\bf k}s)&=&\frac{1}{(2\pi)^{4}}\,\frac{e^{2}\,g^{2}}{8}\,\theta^{2}\int d^{4}l\,\, \,\bar{v}_{r}({\bf q})\gamma^{\alpha}\,(1+\gamma^{5})\,S_{F}(-l)\slashed{\epsilon}_{s^{'}}({\bf k^{'}})
S_{F}(k^{'}-l)\nonumber\\
&&\,\slashed{\epsilon}_{s}({\bf k})\,S_{F}(k^{'}-k-l)\gamma ^{\beta}\,(1+\gamma^{5})\,v_{r^{'}}({\bf q^{'}})\,D_{F_{\alpha\beta}}(l-q),
\end{eqnarray}
where $S_{F}$ denotes the fermion propagator, the indices $r,r'$ and $s,s'$ stand for the Sterile neutrino and photon spin states, respectively.
Now, in order to calculate the forward scattering term in (\ref{forward}), one should find the commutator $[H_I^0(t),D_{ij}^0({\bf p})]$, then evaluate the expectation value $\langle[H_I^0(t),D_{ij}^0({\bf p})]\rangle$  according to the following operator expectation value
\begin{equation}
\langle \, b^\dag_{r'_{i}}(q')b_{r_{j}}(q)\, \rangle
=(2\pi)^3\delta^3(\mathbf{q}-\mathbf{q'})\delta_{rr'}\delta_{ij}\frac{1}{2}f_\text{\tiny{DM}}(\mathbf{x},\mathbf{q}).
\end{equation}
In this regard, one can substitute (\ref{m}-\ref{m4}) into (\ref{H1}) and then (\ref{forward}) to find the time evolution of the density matrix components which is obtained as follows
\begin{eqnarray}\label{rhodott}
\frac{d}{dt}\rho_{ij}(k)\,&=&\,-\frac{\sqrt{2}}{12\,\pi\,k^{0}}\alpha\,\theta^{2}\,{G}_{\text{\tiny{F}}}\int d{\bf{q}}\,\, (\delta_{is}\rho_{s'j}(k)-\delta_{js'}\rho_{is}(k))\,f_\text{\tiny{DM}}({\bf{x}},{\bf{q}})\,
\bar{u}_{r}(q)\,\,(1-\gamma^{5})\nonumber\\&&(q\cdot\epsilon_{s}\,\,\slashed{\epsilon}_{s^{'}}\,+\,q\cdot\epsilon_{s^{'}}\,\,\slashed{\epsilon}_{s})\,u_{r}(q)+\frac{\sqrt{2}}{24\,\pi\,k^{0}}\alpha\,\theta^{2}\,{G}_{\text{\tiny{F}}}\int d{\bf{q}}\,\, (\delta_{is}\rho_{s'j}(k)-\delta_{js'}\rho_{is}(k))\,\nonumber\\&&f_\text{\tiny{DM}}({\bf{x}},{\bf{q}})\,    \bar{u}_{r}(q)\,(1-\gamma^{5})\,\slashed{k}\,(\slashed{\epsilon}_{s^{'}}\,\slashed{\epsilon}_{s}\,-\,\slashed{\epsilon}_{s}\,\slashed{\epsilon_{s^{'}}})\,u_{r}(q).
\end{eqnarray}
}

\color{black}
  

\begin{thebibliography}{99}
  		  	 \bibitem{rotation}
  	 M.S. Roberts and R.N. Whitehurst, "The rotation curve and geometry of M31 at large galactocentric distances", Astrophys. J. \textbf{201}, 327 (1975).
  	\bibitem{White:1977jf}
  	S.~D.~M.~White and M.~J.~Rees,
  	``Core condensation in heavy halos: A Two stage theory for galaxy formation and clusters,''
  	Mon. Not. Roy. Astron. Soc. \textbf{183}, 341-358 (1978)
  	\bibitem{WMAP:2003elm}
  	D.~N.~Spergel \textit{et al.} [WMAP],
  	``First year Wilkinson Microwave Anisotropy Probe (WMAP) observations: Determination of cosmological parameters,''
  	Astrophys. J. Suppl. \textbf{148}, 175-194 (2003)
 \bibitem{Clowe:2003tk}
 D.~Clowe, A.~Gonzalez and M.~Markevitch,
 ``Weak lensing mass reconstruction of the interacting cluster 1E0657-558: Direct evidence for the existence of dark matter,''
 Astrophys. J. \textbf{604}, 596-603 (2004)
 	\bibitem{Cai:2017mow}
 	Y.~Cai, T.~Han, T.~Li and R.~Ruiz,
 	``Lepton Number Violation: Seesaw Models and Their Collider Tests,''
 	Front. in Phys. \textbf{6}, 40 (2018)
 	\bibitem{Baer:2012uy}
 	H.~Baer, V.~Barger, P.~Huang and X.~Tata,
 	``Natural Supersymmetry: LHC, dark matter and ILC searches,''
 	JHEP \textbf{05}, 109 (2012)
 	\bibitem{Sigurdson:2004zp}
 	K.~Sigurdson, M.~Doran, A.~Kurylov, R.~R.~Caldwell and M.~Kamionkowski,
 	``Dark-matter electric and magnetic dipole moments,''
 	Phys. Rev. D \textbf{70}, 083501 (2004)
 	[erratum: Phys. Rev. D \textbf{73}, 089903 (2006)]
 	\bibitem{Kumar:2015wya}
 	J.~Kumar, D.~Marfatia and D.~Yaylali,
 	``Vector dark matter at the LHC,''
 	Phys. Rev. D \textbf{92}, no.9, 095027 (2015)
 	\bibitem{Catena:2018uae}
 	R.~Catena, K.~Fridell and V.~Zema,
 	``Direct detection of fermionic and vector dark matter with polarised targets,''
 	JCAP \textbf{11}, 018 (2018)
 	\bibitem{Graham:2013gfa}
 	P.~W.~Graham and S.~Rajendran,
 	``New Observables for Direct Detection of Axion Dark Matter,''
 	Phys. Rev. D \textbf{88}, 035023 (2013)
 	\bibitem{Duerr:2015aka}
 	M.~Duerr, P.~Fileviez P\'erez and J.~Smirnov,
 	``Scalar Dark Matter: Direct vs. Indirect Detection,''
 	JHEP \textbf{06}, 152 (2016)
 	\bibitem{Arguelles:2012cf}
 	C.~A.~Arguelles and J.~Kopp,
 	``Sterile neutrinos and indirect dark matter searches in IceCube,''
 	JCAP \textbf{07}, 016 (2012)
 	\bibitem{Foster:2021fxa}
 	J.~W.~Foster,
 	``Direct and Indirect Searches for Axion Dark Matter,''
 	\bibitem{Ramos:2021omo}
 	R.~Ramos, Van Que Tran and T.~C.~Yuan,
 	Phys. Rev. D \textbf{103}, no.7, 075021 (2021)
 	\bibitem{Mahmoudi:2018lll}
 	S.~Mahmoudi, M.~Haghighat, S.~A.~M.~Vamegh and R.~Mohammadi,
 	``Dipolar dark matter and CMB B-mode polarization,''
 	Eur. Phys. J. C \textbf{80}, no.5, 402 (2020)
 	\bibitem{ModaresVamegh:2019nja}
 	S.~Modares Vamegh, M.~Haghighat, S.~Mahmoudi and R.~Mohammadi,
 	``Impact of the vector dark matter on polarization of the CMB photon,''
 	Phys. Rev. D \textbf{100}, no.10, 103024 (2019)
 	\bibitem{Haghighat:2019rht}
 	M.~Haghighat, S.~Mahmoudi, R.~Mohammadi, S.~Tizchang and S.~S.~Xue,
 	``Circular polarization of cosmic photons due to their interactions with Sterile neutrino dark matter,''
 	Phys. Rev. D \textbf{101}, no.12, 123016 (2020)
   \bibitem{Green:2021jrr}
   A.~M.~Green,
   ``Dark matter in astrophysics/cosmology,''
   SciPost Phys. Lect. Notes \textbf{37}, 1 (2022)
	\bibitem{Cooley:2021rws}
	J.~Cooley,
	``Dark Matter Direct Detection of Classical WIMPs,''
	[arXiv:2110.02359 [hep-ph]].
	\bibitem{Slatyer:2021qgc}
	T.~R.~Slatyer,
	``Les Houches Lectures on Indirect Detection of Dark Matter,''
	[arXiv:2109.02696 [hep-ph]].
	\bibitem{Arvanitaki:2014wva}
	A.~Arvanitaki, M.~Baryakhtar and X.~Huang,
	``Discovering the QCD Axion with Black Holes and Gravitational Waves,''
	Phys. Rev. D \textbf{91}, no.8, 084011 (2015)
	\bibitem{Cardoso:2018tly}
	V.~Cardoso, \'O.~J.~C.~Dias, G.~S.~Hartnett, M.~Middleton, P.~Pani and J.~E.~Santos,
	``Constraining the mass of dark photons and axion-like particles through black-hole superradiance,''
	JCAP \textbf{03}, 043 (2018)
	\bibitem{Stott:2018opm}
	M.~J.~Stott and D.~J.~E.~Marsh,
	``Black hole spin constraints on the mass spectrum and number of axionlike fields,''
	Phys. Rev. D \textbf{98}, no.8, 083006 (2018)
	
	\bibitem{Hofmann:2019ihc}
	F.~Hofmann and C.~Wegg,
	``7.1 keV sterile neutrino dark matter constraints from a deep Chandra X-ray observation of the Galactic bulge Limiting Window,''
	Astron. Astrophys. \textbf{625}, L7 (2019)
	\bibitem{Chiang:2014xra}
	C.~W.~Chiang and T.~Yamada,
	``3.5-keV X-ray line from nearly-degenerate WIMP dark matter decays,''
	JHEP \textbf{09}, 006 (2014)
	\bibitem{Cadamuro:2011fd}
	D.~Cadamuro and J.~Redondo,
	``Cosmological bounds on pseudo Nambu-Goldstone bosons,''
	JCAP \textbf{02}, 032 (2012)
	\bibitem{Arias:2012az}
	P.~Arias, D.~Cadamuro, M.~Goodsell, J.~Jaeckel, J.~Redondo and A.~Ringwald,
	``WISPy Cold Dark Matter,''
	JCAP \textbf{06}, 013 (2012)
	\bibitem{Higaki:2014qua}
	T.~Higaki, N.~Kitajima and F.~Takahashi,
	``Hidden axion dark matter decaying through mixing with QCD axion and the 3.5 keV X-ray line,''
	JCAP \textbf{12}, 004 (2014)
	\bibitem{Pospelov:2008gg}
	M.~Pospelov, A.~Ritz, C.~Skordis, A.~Ritz and C.~Skordis,
	``Pseudoscalar perturbations and polarization of the cosmic microwave background,''
	Phys. Rev. Lett. \textbf{103}, 051302 (2009)
	\bibitem{Fedderke:2019ajk}
	M.~A.~Fedderke, P.~W.~Graham and S.~Rajendran,
	``Axion Dark Matter Detection with CMB Polarization,''
	Phys. Rev. D \textbf{100}, no.1, 015040 (2019)
	\bibitem{Agrawal:2019lkr}
	P.~Agrawal, A.~Hook and J.~Huang,
	``A CMB Millikan experiment with cosmic axiverse strings,''
	JHEP \textbf{07}, 138 (2020)
	\bibitem{Lue:1998mq}
	A.~Lue, L.~M.~Wang and M.~Kamionkowski,
	``Cosmological signature of new parity violating interactions,''
	Phys. Rev. Lett. \textbf{83}, 1506-1509 (1999)
	\bibitem{Fujita:2020ecn}
	T.~Fujita, K.~Murai, H.~Nakatsuka and S.~Tsujikawa,
	``Detection of isotropic cosmic birefringence and its implications for axionlike particles including dark energy,''
	Phys. Rev. D \textbf{103}, no.4, 043509 (2021)
\bibitem{Liu:2006uh}
G.~C.~Liu, S.~Lee and K.~W.~Ng,
``Effect on cosmic microwave background polarization of coupling of quintessence to pseudoscalar formed from the electromagnetic field and its dual,''
Phys. Rev. Lett. \textbf{97}, 161303 (2006)

	\bibitem{Gasparotto:2022uqo}
	S.~Gasparotto and I.~Obata,
	``Cosmic Birefringence from Monodromic Axion Dark Energy,''
	[arXiv:2203.09409 [astro-ph.CO]].
	\bibitem{Minami:2020odp}
	Y.~Minami and E.~Komatsu,
	``New Extraction of the Cosmic Birefringence from the Planck 2018 Polarization Data,''
	Phys. Rev. Lett. \textbf{125}, no.22, 221301 (2020)
	\bibitem{WMAP:2008lyn}
	E.~Komatsu \textit{et al.} [WMAP],
	``Five-Year Wilkinson Microwave Anisotropy Probe (WMAP) Observations: Cosmological Interpretation,''
	Astrophys. J. Suppl. \textbf{180}, 330-376 (2009)
	\bibitem{Planck:2015qep}
	P.~A.~R.~Ade \textit{et al.} [Planck],
	``Planck 2015 results - II. Low Frequency Instrument data processings,''
	Astron. Astrophys. \textbf{594}, A2 (2016)
	\bibitem{Namikawa:2020ffr}
	T.~Namikawa, Y.~Guan, O.~Darwish, B.~D.~Sherwin, S.~Aiola, N.~Battaglia, J.~A.~Beall, D.~T.~Becker, J.~R.~Bond and E.~Calabrese, \textit{et al.}
	``Atacama Cosmology Telescope: Constraints on cosmic birefringence,''
	Phys. Rev. D \textbf{101}, no.8, 083527 (2020)
	\bibitem{ACT:2020frw}
	S.~K.~Choi \textit{et al.} [ACT],
	``The Atacama Cosmology Telescope: a measurement of the Cosmic Microwave Background power spectra at 98 and 150 GHz,''
	JCAP \textbf{12}, 045 (2020)
	\bibitem{Komatsu:2022nvu}
	E.~Komatsu,
	``New physics from the polarized light of the cosmic microwave background,''
	Nature Rev. Phys. \textbf{4}, no.7, 452-469 (2022)
	\bibitem{Minami:2020fin}
	Y.~Minami and E.~Komatsu,
	``Simultaneous determination of the cosmic birefringence and miscalibrated polarization angles II: Including cross frequency spectra,''
	PTEP \textbf{2020}, no.10, 103E02 (2020)
	\bibitem{Abghari:2022bet}
	A.~Abghari, R.~M.~Sullivan, L.~T.~Hergt and D.~Scott,
	``Constraints on cosmic birefringence using $E$-mode polarisation,''
	[arXiv:2203.10733 [astro-ph.CO]].
	\bibitem{Mohammadi:2021xoh}
	R.~Mohammadi, J.~Khodagholizadeh, M.~Sadegh and A.~Vahedi,
	``Cross-correlation Power Spectra and Cosmic Birefringence of the CMB via Photon-neutrino Interaction,''
	[arXiv:2109.00152 [hep-ph]].
		\bibitem{Khodagholizadeh:2014nfa}
	J.~Khodagholizadeh, R.~Mohammadi and S.~S.~Xue,
	``Photon-neutrino scattering and the B-mode spectrum of CMB photons,''
	Phys. Rev. D \textbf{90} (2014) no.9, 091301
	[arXiv:1406.6213 [astro-ph.CO]].
	\bibitem{Mohammadi:2016bxl}
	R.~Mohammadi, J.~Khodagholizadeh, M.~Sadegh and S.~S.~Xue,
	``B-mode polarization of the CMB and the cosmic neutrino background,''
	Phys. Rev. D \textbf{93} (2016) no.12, 125029
	[arXiv:1602.00237 [astro-ph.CO]].
	\bibitem{Khodagholizadeh:2019het}
	J.~Khodagholizadeh, R.~Mohammadi, M.~Sadegh and A.~Vahedi,
	``B-mode Power Spectrum of CMB via Polarized Compton Scattering,''
	JCAP \textbf{01} (2020), 051
	[arXiv:1909.00568 [astro-ph.CO]].
	\bibitem{Tizchang:2016vef}
	S.~Tizchang, S.~Batebi, M.~Haghighat and R.~Mohammadi,
	``Cosmic microwave background polarization in non-commutative space-time,''
	Eur. Phys. J. C \textbf{76} (2016) no.9, 478
	[arXiv:1605.09045 [hep-ph]].
	\bibitem{2022dipole}
	J.~Khodagholizadeh, R.~Mohammadi, S.~M.~S. Movahed,
	`` CMB Polarization by the Asyemmetric Template of Scalar Perturbation,''
	 Eur. Phys. J. C \textbf{83}, 651 (2023).
	\bibitem{kosowsky1996cosmic}
	A.~Kosowsky,
	``Cosmic microwave background polarization,''
	Annals Phys. \textbf{246} (1996), 49-85
	\bibitem{Bond:1984fp}
	J.~R.~Bond and G.~Efstathiou,
	``Cosmic background radiation anisotropies in universes dominated by nonbaryonic dark matter,''
	Astrophys. J. Lett. \textbf{285} (1984), L45-L48
	\bibitem{Zaldarriaga:1996xe} 
	M.~Zaldarriaga and U.~Seljak,
	Phys.\ Rev.\ D {\bf 55}, 1830 (1997)
		\bibitem{Diego-Palazuelos:2022dsq}
		P.~Diego-Palazuelos, J.~R.~Eskilt, Y.~Minami, M.~Tristram, R.~M.~Sullivan, A.~J.~Banday, R.~B.~Barreiro, H.~K.~Eriksen, K.~M.~G\'orski and R.~Keskitalo, \textit{et al.}
		``Cosmic Birefringence from the Planck Data Release 4,''
		Phys. Rev. Lett. \textbf{128}, no.9, 091302 (2022)
	
	\bibitem{Seljak:1996is} 
	U.~Seljak and M.~Zaldarriaga,
	``A Line of sight integration approach to cosmic microwave background anisotropies,''
	Astrophys.\ J.\  {\bf 469}, 437 (1996)
	

	\bibitem{Masso:2009mu}
	E.~Masso, S.~Mohanty and S.~Rao,
	``Dipolar Dark Matter,''
	Phys. Rev. D \textbf{80}, 036009 (2009)
		\bibitem{Lee:2014koa}
		H.~M.~Lee,
		``Magnetic dark matter for the X-ray line at 3.55 keV,''
		Phys. Lett. B \textbf{738}, 118-122 (2014)
		\bibitem{Davidson:2005cs}
		S.~Davidson, M.~Gorbahn and A.~Santamaria,
		``From transition magnetic moments to majorana neutrino masses,''
		Phys. Lett. B \textbf{626}, 151-160 (2005)
		\bibitem{Kosowsky:1994cy} 
		A.~Kosowsky,
		``Cosmic microwave background polarization,''
		Annals Phys.\  {\bf 246}, 49 (1996)
		\bibitem{Dodelson:2003ft}
		S.~Dodelson,
		``Modern Cosmology,''
		Academic Press, 2003,
		ISBN 978-0-12-219141-1
		
		
		\bibitem{Planck:2018vyg}
		N.~Aghanim \textit{et al.} [Planck],
		``Planck 2018 results. VI. Cosmological parameters,''
		Astron. Astrophys. \textbf{641}, A6 (2020)
		[erratum: Astron. Astrophys. \textbf{652}, C4 (2021)]
		
		\bibitem{Burles:1999zt} 
		S.~Burles, K.~M.~Nollett, J.~W.~Truran and M.~S.~Turner,
		``Sharpening the predictions of big bang nucleosynthesis,''
		Phys.\ Rev.\ Lett.\  {\bf 82}, 4176 (1999)
		
		
		
		
		
		\bibitem{Boyarsky:2018tvu} 
		A.~Boyarsky, M.~Drewes, T.~Lasserre, S.~Mertens and O.~Ruchayskiy,
		``Sterile neutrino Dark Matter,''
		Prog.\ Part.\ Nucl.\ Phys.\  {\bf 104}, 1 (2019)
			\bibitem{Dolgov:2002wy} 
			A.~D.~Dolgov,
			``Neutrinos in cosmology,''
			Phys.\ Rept.\  {\bf 370}, 333 (2002)
			\bibitem{Bezrukov:2017ike} 
			F.~Bezrukov, A.~Chudaykin and D.~Gorbunov,
			``Hiding an elephant: heavy Sterile neutrino with large mixing angle does not contradict cosmology,''
			JCAP {\bf 1706}, no. 06, 051 (2017)
\bibitem{Dodelson:1993je}
S.~Dodelson and L.~M.~Widrow,
``Sterile-neutrinos as dark matter,''
Phys. Rev. Lett. \textbf{72}, 17-20 (1994)

		\bibitem{Buckley:2013bha}
		J.~Buckley, D.~F.~Cowen, S.~Profumo, A.~Archer, M.~Cahill-Rowley, R.~Cotta, S.~Digel, A.~Drlica-Wagner, F.~Ferrer and S.~Funk, \textit{et al.}
		``Working Group Report: WIMP Dark Matter Indirect Detection,''
		[arXiv:1310.7040 [astro-ph.HE]].
					\bibitem{Boyarsky:2005us}
					A.~Boyarsky, A.~Neronov, O.~Ruchayskiy and M.~Shaposhnikov,
					``Constraints on sterile neutrino as a dark matter candidate from the diffuse x-ray background,''
					Mon. Not. Roy. Astron. Soc. \textbf{370}, 213-218 (2006).
					\bibitem{Boyarsky:2006fg}
					A.~Boyarsky, A.~Neronov, O.~Ruchayskiy, M.~Shaposhnikov and I.~Tkachev,
					``Where to find a dark matter sterile neutrino?,''
					Phys. Rev. Lett. \textbf{97}, 261302 (2006).
					\bibitem{Boyarsky:2006ag}
					A.~Boyarsky, J.~Nevalainen and O.~Ruchayskiy,
					``Constraints on the parameters of radiatively decaying dark matter from the dark matter halo of the Milky Way and Ursa Minor,''
					Astron. Astrophys. \textbf{471}, 51-57 (2007).
					\bibitem{Boyarsky:2007ay}
					A.~Boyarsky, D.~Iakubovskyi, O.~Ruchayskiy and V.~Savchenko,
					``Constraints on decaying Dark Matter from XMM-Newton observations of M31,''
					Mon. Not. Roy. Astron. Soc. \textbf{387}, 1361 (2008).
					\bibitem{Yuksel:2007xh}
					H.~Yuksel, J.~F.~Beacom and C.~R.~Watson,
					``Strong Upper Limits on Sterile Neutrino Warm Dark Matter,''
					Phys. Rev. Lett. \textbf{101}, 121301 (2008).
					\bibitem{Seljak:2006qw}
					U.~Seljak, A.~Makarov, P.~McDonald and H.~Trac,
					``Can sterile neutrinos be the dark matter?,''
					Phys. Rev. Lett. \textbf{97}, 191303 (2006).
					\bibitem{Boyarsky:2008xj}
					A.~Boyarsky, J.~Lesgourgues, O.~Ruchayskiy and M.~Viel,
					``Lyman-alpha constraints on warm and on warm-plus-cold dark matter models,''
					JCAP \textbf{05}, 012 (2009).
					\bibitem{Perez:2016tcq}
					K.~Perez, K.~C.~Y.~Ng, J.~F.~Beacom, C.~Hersh, S.~Horiuchi and R.~Krivonos,
					``Almost closing the $\ensuremath{\nu}$MSM sterile neutrino dark matter window with NuSTAR,''
					Phys. Rev. D \textbf{95}, no.12, 123002 (2017).
					\bibitem{Shi:1998km}
					X.~D.~Shi and G.~M.~Fuller,
					``A New dark matter candidate: Nonthermal sterile neutrinos,''
					Phys. Rev. Lett. \textbf{82}, 2832-2835 (1999)
					\bibitem{Shaposhnikov:2006xi}
					M.~Shaposhnikov and I.~Tkachev,
					``The nuMSM, inflation, and dark matter,''
					Phys. Lett. B \textbf{639}, 414-417 (2006).
					\bibitem{Kusenko:2006rh}
					A.~Kusenko,
					``Sterile neutrinos, dark matter, and the pulsar velocities in models with a Higgs singlet,''
					Phys. Rev. Lett. \textbf{97}, 241301 (2006).
					\bibitem{Khalil:2008kp}
					S.~Khalil and O.~Seto,
					``Sterile neutrino dark matter in B - L extension of the standard model and galactic 511-keV line,''
					JCAP \textbf{10}, 024 (2008).
					\bibitem{Kaneta:2016vkq}
					K.~Kaneta, Z.~Kang and H.~S.~Lee,
					``Right-handed neutrino dark matter under the B$-$L gauge interaction,''
					JHEP \textbf{02}, 031 (2017).
					\bibitem{Biswas:2016bfo}
					A.~Biswas and A.~Gupta,
					``Freeze-in Production of Sterile Neutrino Dark Matter in U(1)$_{\rm B-L}$ Model,''
					JCAP \textbf{09}, 044 (2016).
					\bibitem{Seto:2020udg}
					O.~Seto and T.~Shimomura,
					``Signal from sterile neutrino dark matter in extra $U(1)$ model at direct detection experiment,''
					Phys. Lett. B \textbf{811}, 135880 (2020).
			\bibitem{Xue:2016dpl}
			S.~S.~Xue,
			``Hierarchy spectrum of SM fermions: from top quark to electron neutrino,''
			JHEP {\bf 1611}, 072 (2016)
			\bibitem{Xue:1997tz} 
			S.~S.~Xue,
			``Neutrino masses and mixings,''
			Mod.\ Phys.\ Lett.\ A {\bf 14}, 2701 (1999)
				\bibitem{Xue:2015wha} 
				S.~S.~Xue,
				``Vectorlike $W^\pm$-boson coupling at TeV and third family fermion masses,''
				Phys.\ Rev.\ D {\bf 93}, no. 7, 073001 (2016)
				\bibitem{Xue:1996fm} 
				S.~S.~Xue,
				``Quark masses and mixing angles,''
				Phys.\ Lett.\ B {\bf 398}, 177 (1997)
\end{thebibliography}
\end{document}